\documentclass[superscriptaddress,aps,prd,showpacs,twocolumn]{revtex4}
\usepackage{amssymb}

\usepackage{eurosym}
\usepackage{graphicx}
\usepackage{amsmath}
\usepackage{mathrsfs}
\usepackage{bm}
\usepackage{color}

\def\be{\begin{equation}}
\def\ee{\end{equation}}
\def\bea{\begin{eqnarray}}
\def\eea{\end{eqnarray}}

\begin{document}

\title{The minimum mass of a charged spherically symmetric object in $D$ dimensions,
its implications for fundamental particles, and holography}
\author{Piyabut Burikham}
\email{piyabut@gmail.com}
\affiliation{High Energy Physics Theory Group, Department of Physics, Faculty of Science,
Chulalongkorn University, Phyathai Rd., Bangkok 10330, Thailand,}
\author{Krai Cheamsawat}
\email{kraicwt@gmail.com}
\affiliation{High Energy Physics Theory Group, Department of Physics, Faculty of Science,
Chulalongkorn University, Phyathai Rd., Bangkok 10330, Thailand,}
\affiliation{Theoretical Physics Group, Blackett Laboratory, Imperial College, London SW7 2AZ, United Kingdom,}
\author{Tiberiu Harko}
\email{t.harko@ucl.ac.uk}
\affiliation{Department of Physics, Babes-Bolyai University, Kogalniceanu
Street, Cluj-Napoca 400084, Romania,}
\affiliation{Department of Mathematics, University College London, Gower Street, London
WC1E 6BT, United Kingdom.}
\author{Matthew J. Lake}
\email{matthewj@nu.ac.th}
\affiliation{The Institute for Fundamental Study, ``The Tah Poe Academia Institute", \\
Naresuan University, Phitsanulok 65000, Thailand and \\
Thailand Center of Excellence in Physics, Ministry of Education, Bangkok
10400, Thailand. }
\date{\today }

\begin{abstract}
We obtain bounds for the minimum and maximum mass/radius ratio of a stable,
charged, spherically symmetric compact object in a $D-$dimensional
space-time in the framework of general relativity, and in the presence of dark energy. The total energy, including the gravitational
component, and the stability of objects with minimum mass/radius ratio
is also investigated. The minimum energy condition leads to a representation of the mass and radius of the
charged objects with minimum mass/radius ratio in terms of the charge and
vacuum energy only. As applied to the electron in the four-dimensional case, this procedure allows one to re-obtain the classical electron radius from purely general relativistic considerations. By combining the lower mass bound, in four space-time dimensions, with
minimum length uncertainty relations (MLUR) motivated by quantum gravity, we
obtain an alternative bound for the maximum charge/mass ratio of a stable, gravitating, charged quantum mechanical object, expressed in terms of fundamental constants. Evaluating
this limit numerically, we obtain again the correct order of magnitude value for
the charge/mass ratio of the electron, as required by the stability conditions. This
suggests that, if the electron were either less massive (with the same
charge) or if its charge were any higher (for fixed mass), a
combination of electrostatic and dark energy repulsion would destabilize the
Compton radius. In other words, the electron would blow itself apart. Our results suggest the existence of a deep connection between
gravity, the presence of the cosmological constant, and the stability of
fundamental particles.

\vspace{5mm}

{\bf Keywords}: general relativity; higher dimensional gravity; charged objects; mass bounds; minimum length uncertainty relation; holography
\end{abstract}
\pacs{04.20.Cv; 04.50.-h; 04.60.Bc; 04.40.Nr}

\maketitle


\section{Introduction} \label{Sec.1}

The existence of a minimum length is an important prediction of
phenomenological quantum gravity. A fundamental bound yielding the smallest
resolvable length scale could help solve several outstanding problems in
theoretical physics, for example, by providing a natural cut off to
regularize divergent integrals in the renormalization of quantum field
theories, or by preventing matter from collapsing to form a singularity at
the centre of a black hole. Furthermore, the existence of both minimum and
maximum length scales in nature or, at least, at a given epoch (for example,
$R_U \approx 1.3 \times 10^{28}$ cm is the current size of the horizon and
acts as a {\it de facto} maximum length scale for physical phenomena in the
Universe today), is naturally linked to the existence of upper and lower
bounds on the mass-energy scales of physical processes. In this paper, we
determine bounds on the mass/radius ratio of stable charged objects, both
classically and quantum mechanically, and investigate their implications for
fundamental particles.

One way to introduce a minimum length is via a Generalized Uncertainty
Principle (GUP) that extends the usual Heisenberg Uncertainty Principle
(HUP) to include nonlinear terms, which may then be interpreted as quantum
gravity effects. A GUP of the form
\begin{eqnarray}  \label{GUP}
\Delta x \Delta p\geq \frac{\hbar}{2}\left[1+\mathcal{A} \left(\Delta
p\right)^2+\mathcal{B} \right],
\end{eqnarray}
where $\mathcal{A}$ and $\mathcal{B}$ are positive constants, was proposed in \cite{Kempf}%
, and many different modifications of the HUP have since been considered in
the literature. The GUP in Eqn.~(\ref{GUP}) gives rise to an effective minimum length,
in the form of a minimum positional uncertainty,
\begin{eqnarray}
\Delta x_{\rm min}&=& \hbar\sqrt{\mathcal{A}(1+\mathcal{B})},
\end{eqnarray}
which is proportional to $\sqrt{\mathcal{A}}$, but the existence of a minimum bound
is a general feature of these models. Collectively, such modified relations are
referred to either as GUPs, or as Minimum Length Uncertainty Relations
(MLURs) (for general reviews of
GUP phenomenology see \cite{Tawfik:2014zca,Tawfik:2015rva}, and see \cite{Garay:1994en,Hossenfelder:2012jw} for reviews of
minimum length scenarios in quantum gravity). The existence of an absolute bound
of the form $\Delta x \geq \Delta x_{min}$ implies that $\Delta x$ cannot be made arbitrarily small,
irrespective of the uncertainties in any other physical observables.

It is interesting to note that the idea of a minimum length induced by
quantum gravitational effects was first proposed long ago \cite{Bronstein}.
By investigating the quantum mechanical measurement of the $%
\Gamma^{0}_{01}$ component of the Christoffel symbols, Bronstein obtained a
fundamental limit for the temporal uncertainty inherent in the measurement
process,
\begin{eqnarray}  \label{Bronstein-1}
\Delta t \gtrsim \left(\frac{\hbar}{c^2G\rho^2V}\right)^{1/3},
\end{eqnarray}
where $\rho$ and $V$ denote the density and the volume of a massive body, respectively.
This, in turn, may be related to the spatial
uncertainty via $\Delta x \leq c \Delta t$. By taking into account that $M =
\rho V$ is the particle mass, we obtain an equivalent mass$-$time$-$density
uncertainty relation of the form
\begin{eqnarray}  \label{Bronstein-2}
M \gtrsim \frac{\hbar}{c^2G \rho \left(\Delta t \right)^3}.
\end{eqnarray}

Since the existence of lower bounds for physical quantities is a natural
characteristic of quantum processes, the presence of similar bounds in the
framework of classical physics appears, at first sight, somehow unusual.
Nonetheless, lower bounds on the ratios of physical quantities do occur
naturally in classical general relativity, as a form of stability condition
for compact objects. Two such bounds are of particular interest for both
astrophysics/cosmology and for the study of subatomic particles: the minimum
mass/radius ratio for a compact object in the presence of dark energy and
for a charged compact object.

Classical $(3+1)$-dimensional general relativity, with no dark energy
component ($\Lambda = 0$), imposes an upper bound on the mass/radius ratio
of any compact object, the Buchdahl limit \cite{Buch}, which requires a
sphere of matter with arbitrary equation of state to satisfy the stability
constraint
\begin{eqnarray}  \label{Buch}
\frac{2GM}{c^2R} \leq \frac{8}{9}.
\end{eqnarray}
If this condition is violated, the object will inevitably collapse under its
own gravity to form a black hole. (Typically, this process occurs for stars
when the mass of the star exceeds approximately $3.2M_{\odot}$ \cite{Ruf}.)
The Buchdahl limit, and its extensions, have been intensively investigated \cite{Arnowitt:1960zza,Arnowitt:1960zzb}, 
including the study of the effects of the cosmological constant \cite{Har1},
and of sharp limits on the mass/radius bounds \cite{Andr1, Andr2,Andr3}.
$D$-dimensional extensions of the Buchdahl limits in the presence of a
cosmological constant were obtained in \cite{Zarro, Wright1}, while the
mass/radius ratio for compact objects in five dimensional Gauss-Bonnet
gravity and $f(R)$ gravity were considered in \cite{Wright2} and \cite%
{Goswami}, respectively.

In the presence of dark energy, a minimum bound for the mass/radius ratio of
a stable compact object also exists. This result follows rigorously from the
generalized Buchdahl inequalities for a compact object in the presence of a
nonzero cosmological constant ($\Lambda \neq 0$) \cite{Boehmer:2005sm}. For $%
\Lambda > 0$, the existence of a lower bound admits an intuitive
explanation: If the stability condition is violated, the self gravity of the
object is insufficient to overcome the repulsive force due to dark energy. %
Remarkably, a minimum mass also exists for $\Lambda < 0$
\cite{Boehmer:2005sm}. Physically, this is due to the balancing of both gravitational and dark energy
attraction with local pressure in the matter distribution, induced by non-gravitational forces.

In \cite{Boehmer:2005sm}, it was shown that a
compact object is stable against dark energy repulsion when its density is
above a certain minimum value,
\begin{eqnarray}  \label{min_dens}
\rho \geq \rho _{\Lambda} = \frac{\Lambda c^2}{16\pi G},
\end{eqnarray}
for $\Lambda > 0$. A similar condition follows from the generalized Buchdahl inequality for a
charged compact object, even in the absence of dark energy~\cite{Boehmer:2007gq}. For $\Lambda = 0$, this gives
\begin{eqnarray}  \label{eqn:b}
\frac{2GM}{c^2 R}\geq \frac{3}{2}\frac{GQ^{2}}{c^4 R^{2}} \left[\frac{1+%
\frac{GQ^2}{18c^4R^2}}{1+ \frac{GQ^{2}}{12c^4R^{2}}}\right].
\end{eqnarray}
 For $\Lambda \neq 0$, this result generalizes to \cite{Boehmer:2007gq}
\begin{eqnarray}  \label{eqn:a}
\frac{2GM}{c^2 R}\geq \frac{3}{2}\frac{GQ^{2}}{c^4R^{2}} \left[\frac{1+%
\frac{1}{9} \frac{c^4\Lambda R^{4}}{GQ^{2}} - \frac{1}{54}\Lambda R^{2}
+\frac{GQ^2}{18c^4R^2}} {1+\frac{GQ^{2}}{12c^4R^{2}}}\right].
\end{eqnarray}
Hence, for $R^2\Lambda \ll 1$, the effect of dark energy is subdominant to
electrostatic repulsion. Equation (\ref{eqn:b}) can also be Taylor expanded to
give
\begin{eqnarray}  \label{eqn:c}
\frac{2GM}{c^2R}\geq \frac{3}{2}\frac{GQ^{2}}{c^4R^{2}} \Bigl(1-\frac{%
GQ^2}{36c^4R^2} + O(Q^2/R^2)^4 \Bigr),
\end{eqnarray}
so that, to leading order, we have
\begin{eqnarray}  \label{R_class}
R \gtrsim \frac{3}{4}\frac{Q^2}{Mc^2}.
\end{eqnarray}
In this limit, we recover the standard expression for the classical radius
of a charged body with mass $M$ and charge $Q$: that is, the radius at
which the electrostatic potential energy associated with the object is equal to its
rest mass, $Mc^2$. We recall that this is roughly the radius the object would have \emph{if}
its mass were due only to electrostatic potential energy.

Several general restrictions on the total charge $Q$ of a stable compact
object can also be obtained from the study of the behavior of the Ricci
invariants $r_0 = \mathcal{R}_i^i = \mathcal{R}$, $r_1 = \mathcal{R}_{ij}\mathcal{R}^{ij}$ and $r_2 =
\mathcal{R}_{ijkl}\mathcal{R}^{ijkl}$. For example, by considering that the surface density must
vanish, it may be shown that $Q$ satisfies the condition
\begin{eqnarray}  \label{surface}
\frac{Q^2}{R^4} < \frac{\rho _c c^2}{2}\sqrt{1+\left(\frac{p_c }{\rho
_cc^2}\right)^2},
\end{eqnarray}
where $\rho _c$ and $p_c$ are the central density and pressure of the
object, respectively.

Though most investigations of stellar structure have been done under the
assumption of charge neutrality, there are a number of physical processes
that could lead to the formation of charged regions inside compact objects.
One of these processes could be mass accretion by a neutron star \cite{Har2}%
, if it happens that accretion produces luminosities very close to the
Eddington limit $L_E=4\pi GMm_pc/\sigma _T$ \cite{ShTe83}, where $M$ is the
mass of the star, $\sigma _T$ is the Thomson scattering cross section, and $%
m_p$ is the mass of the proton. Let us assume that the star undergoes
spherical accretion, and that the accreting material is ionized hydrogen. If
the accreting luminosity of the star is $L$, then infalling electrons, at a
distance $r$ from the centre of the star, experience a radiative force $%
F_{rad}=\sigma _TL/4\pi cr^2$ \cite{Har2}. On the other hand, the radiation
drag acting on the protons is smaller by a factor $\left(m_e/m_p\right)^2
\approx 3 \times 10^{-7}$, where $m_e$ is the mass of the electron, so that
electrons and protons are subjected to different accelerations. Therefore, a
star can acquire a net positive charge, $Q=\left(GMm_p/e\right)%
\left(L/L_E\right)$, through accretion \cite{Har2}.

Another possibility
giving rise to the existence of charged macroscopic objects is related to
quark deconfinement inside dense neutron matter \cite{Harq}. If
deconfinement occurs inside a dense neutron star, the strange
beta-equilibrated quark matter consists of an approximately equal mixture of
three quarks, the up, down and strange quarks, with a slight deficit in the
number of strange quarks. This composition of quark matter could lead to a
net positive charge inside the neutron star or quark star.

In deriving the results quoted above, it was assumed that the pressure
within the object is isotropic. Interestingly, anisotropies in the pressure
distribution inside compact objects, in the presence of a cosmological
constant, can significantly modify both the upper and lower bounds for the
mass. These bounds are strongly dependent on the anisotropy parameter $\Sigma
$, which is defined as the difference between the tangential and radial
pressure at the surface of the object. Pressure anisotropies modify the
lower bound on the minimum density of a stable spherical mass distribution,
for $\Lambda > 0$, so that \cite{Boehm2}
\begin{eqnarray}
\rho \geq \rho_{\Lambda}(\Sigma) = \frac{\Lambda c^2}{16\pi G}\left[\frac{1
- \frac{G\Sigma R^2}{6c^2}}{1 + \frac{G\Sigma R^2}{12c^2}}\right].
\end{eqnarray}
Hence, the presence of the anisotropic pressure distribution weakens the
lower bound on the mass. Remarkably, even anisotropic objects may still be
stable, as long as their mass exceeds an absolute classical minimum value,
determined by both $\Lambda$ and $\Sigma$. The existence of a cosmological
constant therefore has profound consequences for the stability of matter,
even at the classical level.

The nature of the cosmological constant or, more generally, dark energy, is
one of the most fundamental problems in contemporary physics. In particular,
the important question of whether $\Lambda$ represents a true fundamental
constant of nature, or simply an approximation (for example, an
approximately constant field configuration that arises as a solution to the,
as yet unknown, equations of motion for a dynamical scalar field), remains
unanswered. However, even if we take the existence of the cosmological
constant, as implied by the Cosmological Concordance, or $\Lambda$CDM model
(c.f. \cite%
{Ostriker:1995rn,Tegmark:2003ud,Tegmark:2000qy,Hazra:2014hma,Zunckel:2008ti}%
), at face value, yet another question remains: Is $\Lambda$ an \emph{%
independent} constant of nature, or can it be expressed in terms of other,
known constants of nature?

In \cite{Boehm3}, it was shown that, if the minimum mass in nature is $M_W = (\hbar/c)\sqrt{\Lambda/3}
\approx 3.5 \times 10^{-66}$ g, as proposed by Wesson \cite{Wesson:2003qn}, then a particle with mass $M_W$ and density $%
\rho_{\Lambda}$, given by Eq. (\ref{min_dens}), has a \emph{classical}
radius given by
\begin{eqnarray}  \label{R}
R = (R_P^2R_W)^{1/3} \approx 2.967 \times 10^{-13} \ \mathrm{cm},
\end{eqnarray}
where $R_P$ is the Planck length, and $R_W=\sqrt{3/\Lambda } \approx R_U \approx 10^{26}$ m.
This is of the same order of magnitude as the classical electron radius,
\begin{eqnarray}  \label{r_e}
r_e = \frac{e^2}{m_e c^2} \approx 2.812 \times 10^{-13} \ \mathrm{cm}.
\end{eqnarray}
Based on this observation, it was suggested in \cite{Boehm3} that the radius
$R$, given by Eq. (\ref{R}), should be formally identified with $%
r_e$ and taken as a minimum possible length scale in nature. The
cosmological constant $\Lambda$ may then be formally identified with the
`standard' set of physical constants $\left\{c,G,\hbar,e,m_e\right\}$
via
\begin{eqnarray}  \label{Lambda}
\Lambda = 3\frac{m_{e}^6G^2}{\alpha^6\hbar^4},
\end{eqnarray}
where $\alpha = e^2/q_P^2$ is the fine structure constant and $q_P = \sqrt{ \hbar c}$
is the Planck charge. Evaluating this numerically gives $\Lambda = 1.4\times 10^{-56}$ cm$^{-2}$, in good
agreement with the value inferred from various cosmological observations
\cite%
{Ostriker:1995rn,Tegmark:2003ud,Tegmark:2000qy,Hazra:2014hma,Zunckel:2008ti}.
In \cite{Boehm3}, the formal identification $R= r_e$ was
justified on the basis of a `small number hypothesis', which represents an
extension of the large number hypothesis proposed by Dirac \cite{Dirac}, and
which proposes that a numerical equality between two very small quantities
with a very similar physical meaning cannot be a coincidence. Interestingly,
the same identification was also obtained in \cite{Beck:2008rd} using
information theory, in which a set of axioms for the cosmological constant
were formulated by analogy with the Khinchin axioms \cite{Khinchin}, by
formally replacing the dependency of the information measure on
probabilities of events by a dependency of $\Lambda$ on the other
fundamental constants of nature. These results raise the interesting questions of whether there is an
intrinsic relation between electromagnetic phenomena and dark energy, and of
what form the possible interaction/coupling of the electric charge $e$ with
the cosmological constant $\Lambda$ may be.

In this work, we aim to show concretely that, by consistently combining
results from general relativity, canonical quantum theory, and MLURs
predicted by phenomenological quantum gravity, the identification (\ref{Lambda})
can be explicitly obtained by saturation of the quantum
gravitational stability condition for the electron. Furthermore, our results
show this identification to be \emph{broadly} consistent with the results obtained by
various early pioneers of quantum gravity research (see \cite{Garay:1994en,Hossenfelder:2012jw} for reviews),
including those of Bronstein \cite{Bronstein} and those obtained by K{\' a}rolyh{\' a}zy \emph{et al} \cite{Karolyhazy:1966zz,KFL,Diosi:1993vy}.

However, some assumptions present in the existing quantum gravity literature are shown to
be inconsistent with canonical quantum mechanics. Specifically, certain
assumptions about the nature of MLURs imply quantum gravity effects which
manifest on scales larger than the observed Compton wavelengths of
elementary particles. (Clearly, this cannot be the case, otherwise quantum
gravity would already have been observed in the lab.) Interestingly,
when these assumptions are revised in order to ensure the consistency of MLURs
with the canonical theory (i.e. by ensuring that quantum gravity effects are
subdominant to `standard' quantum effects), the results obtained are \emph{inconsistent}
with both Bronstein's formulation, Eqs. (\ref{Bronstein-1})-(\ref{Bronstein-2}), and K{\' a}rolyh{\' a}zy's
original results \cite{Karolyhazy:1966zz,KFL}. The reasons for these discrepancies are discussed in detail in Sec. \ref{Sec.7}.

The structure of this paper is as follows. In Sections~\ref{Sec.2}-\ref{Sec.3},
we obtain the generalized Buchdahl inequalities, in arbitrary space-time
dimensions, for a charged spherically symmetric object embedded in a
space-time with general nonvanishing (i.e. positive or negative) %
dark energy. This extends previous results given in \cite%
{Burikham:2015nma}, in which the $D-$dimensional generalized Buchdahl
inequalities for uncharged matter were obtained in both the %
asymptotically de Sitter and anti-de Sitter cases.
Specifically, in Section~\ref{Sec.2}, the gravitational field equations and the
hydrostatic equilibrium equations, also known as the
Tolman-Oppenheimer-Volkov (TOV) equations, are obtained. The general form of
the mass limits are given in Sec. \ref{Sec.3} and various limiting cases of
special physical interest are considered in Sec. \ref{Sec.4}. In Sec. \ref{Sec.5}, we use our previous
results to derive bounds on the minimum and maximum densities of \emph{static} asymptotically de Sitter and anti-de Sitter space-times.
(These results are interesting because, even if the \emph{real} universe is an expanding $(3+1)$-dimensional space-time with a positive cosmological constant,
these static, asymptotically de Sitter and anti-de Sitter spaces still have essential interpretations from the viewpoint of holographic duality.)
The thermodynamic stability of higher-dimensional charged objects is investigated in Sec.~\ref{Sec.6}.
By minimizing the gravitational energy of charged objects with minimum mass/radius ratio,
we show that the ratio of the square of the charge of the object to its mass, $Q^2/M$, is proportional to the radius of the object, $R$, (to leading order).
In Section~\ref{Sec.7}, we investigate the quantum mechanical implications of the lower
mass bound for charged objects, in the standard $(3+1)$-dimensional
scenario, leading to the identification of the cosmological constant $\Lambda
$ in terms of other fundamental constants of nature, as in Eq. (\ref{Lambda}%
). This identification is seen to arise as a consequence of the stability
bound for the electron, viewed as a charged, gravitating, quantum mechanical
particle, and extends the results obtained in \cite{Burikham:2015nma}, in
which the minimum mass of an uncharged, gravitating, quantum mechanical
particle was determined.  Section \ref{Sec.8} shows that MLUR leads to holography in arbitrary noncompact dimensions, and discusses its relation to the results previously obtained in \ref{Sec.7}.
Section \ref{Sec.9} contains a summary of our main results and a brief discussion of possibilities for future work.


\section{Geometry, field equations, and the TOV equations for charged objects in $D$ dimensions} \label{Sec.2}

In the following, we assume that the line element of a spherically symmetric $%
D$-dimensional static space-time can be represented in a generic form as \cite{D}
\begin{equation}
ds^2 = e^{\nu(r)}c^2dt^2 - e^{\lambda(r)}dr^2 - r^2d\Omega^2_{D-2},
\end{equation}
where
\begin{equation}
d\Omega^2_{D-2}=d\theta _1^2+\sin ^2\theta _1d\theta _2^2+...+\sin^2\theta
_1...\sin ^2\theta _{D-3}d\phi ^2.
\end{equation}
Here $x^0=ct$, $x^1=r$, where $r$ is the radial coordinate in $D$ space-%
time dimensions, with domain $0\leq r <\infty$, while the angular
coordinates are defined according to $0\leq \theta _i\leq \pi$, $i=1,...,D-3$%
, and $0\leq \phi \leq 2\pi$, respectively.

The Einstein gravitational field equations are given by
\begin{equation}
G^{\mu}_{\nu}\equiv R^{\mu}_{\phantom{1}\nu} - \frac{1}{2}\delta^{\mu}_{%
\phantom{1}\nu}R = \kappa\left( T^{\mu(M)}_{\phantom{1}\nu}+T^{\mu(DE)}_{%
\phantom{1}\nu} +T^{\mu(EM)}_{\phantom{1}\nu} \right),
\end{equation}
where $\kappa = 8\pi G_D/c^{4}$ and the energy-momentum tensor contains three components,
corresponding to matter $(M)$, dark energy $(DE)$, and the electromagnetic field $(EM)$.
We also assume that the matter and dark energy parts may be expressed in terms of fluid variables, so that
\begin{eqnarray}
T^{\mu(M)}_{\phantom{1}\nu} &=& (\rho
c^2+P)u^{\mu}u_{\nu}-P\delta^{\mu}_{\nu} , \\
T^{\mu(DE)}_{\phantom{1}\nu} &=& (\rho_{DE}
c^2+P_{DE})u^{\mu}u_{\nu}-P_{DE}\delta^{\mu}_{\nu},
\end{eqnarray}
with the dark energy obeying the equation of state $P_{DE} = w\rho_{DE}c^2$,
where $\rho_{DE} =\Lambda_D c^2/8\pi G_D$, and $w=\mathrm{constant}$.
Finally, the electromagnetic energy-momentum tensor is given by
\begin{eqnarray}
T^{\mu(EM)}_{\phantom{1}\nu} &=& F^{\mu}_{\phantom{1}{\alpha}}F^{\alpha}_{%
\phantom{1}\nu} + \frac{1}{4}\delta^{\mu}_{\nu}F_{\alpha\beta}F^{\alpha%
\beta}
\end{eqnarray}
where $F_{\mu \nu}=\nabla _{\nu} A_{\mu}-\nabla _{\mu}A_{\nu}$ and $%
A_{\mu}$ denotes the electromagnetic vector potential. The electromagnetic field tensor $F_{\mu \nu}$ satisfies the cyclic identity
\begin{equation}  \label{max}
\partial _{\lambda}F_{\mu \nu}+\partial _{\mu}F_{\nu \lambda}+\partial
_{\nu}F_{\lambda \mu }=0,
\end{equation}
and the Maxwell equations
\begin{equation}
\frac{1}{\sqrt{-g}}\partial_{\mu}\left(\sqrt{-g}F^{\mu\nu}\right) = j^{\nu},
\end{equation}
where $j^{\nu}$ denotes the electric current four-vector.
We choose the rest frame of the fluid so that the $D$-velocity is
\begin{equation}
u^{\mu} = \delta^{\mu}_t e^{-\nu/2},
\end{equation}
which is then normalized according to $u^2 = g_{tt}u^tu^t = 1$. We
also introduce the electric charge density $\rho _e$, which is related to the time component of the current four-vector by the relation
\begin{equation}
j^t = \rho_e(r) u^t = \rho_e(r)e^{-\frac{\nu}{2}}.
\end{equation}
 From the Maxwell Eqs.~(\ref%
{max}), and with the use of the charge density defined above, we can construct
the `proper charge', $Q(r)$, defined as
\begin{equation}
Q(r) = \int_0^r\rho_e\left(e^{\frac{\lambda(r^{\prime })}{2}}\right)r^{\prime
D-2}dr^{\prime },
\end{equation}
where we note that the definition does not contain the angular volume $%
\Omega_{D-2}$.

By direct substitution, we obtain the nonzero components of the energy-momentum tensor of the electromagnetic
field  as
\begin{eqnarray}
F^{tr} &=& -\frac{e^{-\frac{\nu+\lambda}{2}}}{r^{D-2}}Q(r) = -F^{rt}, \\
F_{tr} &=& g_{tt}g_{rr}F^{tr} = \frac{e^{\frac{\nu+\lambda}{2}}}{r^{D-2}}%
Q(r) = -F_{rt}, \\
F_t^{\phantom{1}r} &=& g_{tt}F^{tr} = -\frac{e^{\frac{\nu-\lambda}{2}}}{%
r^{D-2}}Q(r), \\
F^t_{\phantom{1}r} &=& g^{tt}F_{tr} = \frac{e^{\frac{-\nu+\lambda}{2}}}{%
r^{D-2}}Q(r), \\
F^r_{\phantom{1}t} &=& g^{rr}F_{rt} = \frac{e^{\frac{\nu-\lambda}{2}}}{%
r^{D-2}}Q(r).
\end{eqnarray}
For later use, we also compute $F^{2} = F_{\alpha\beta}F^{\alpha\beta}$, which contains only contributions from $F_{rt}$ and $F_{tr}$, so that
\begin{eqnarray}
F^2 = F_{tr}F^{tr}+F_{rt}F^{rt} = -2\frac{Q^2(r)
}{r^{2(D-2)}}. \nonumber
\end{eqnarray}
Thus, for the components of the electromagnetic energy-momentum tensor, we obtain
\begin{eqnarray}
T^{t(EM)}_{\phantom{1}t} &=& T^{r(EM)}_{\phantom{1}r}= - T^{\theta_i(EM)}_{\phantom{1}
\theta_i}=\frac{Q^2(r)}{2r^{2(D-2)}},  \label{emT}
\end{eqnarray}
where $i=1,2,...,D-2$ denote the angular variables of a $D-$dimensional space-time.
Hence, the electric field from the charge density generates a positive energy density and a
radial pressure equal to $Q^{2}/2r^{2(D-2)}$, and a negative transverse pressure with exactly the same magnitude.

The conservation of the \textit{total} energy-momentum tensor
\begin{equation}
\nabla_{\mu}T^{\mu}_{\phantom{1}\nu} = 0,
\end{equation}
gives the equation
\begin{eqnarray}
-\partial_rP &-& (\rho c^{2}+P)\frac{\partial_r\nu}{2} -\partial_rP_{DE}-(\rho
c^{2}+P)_{DE}\frac{\partial_r\nu}{2}
\notag \\
&+ &\frac{Q}{r^{2(D-2)}}\partial_rQ(r)=0.
\end{eqnarray}
This can be rewritten in a more compact form as
\begin{equation}
\partial_rP_{eff} = -(\rho c^{2}+\rho_{DE}c^{2}+P_{eff})\frac{\partial_r\nu}{%
2} + \frac{Q}{r^{2(D-2)}}\partial_rQ(r) ,  \label{conservation}
\end{equation}
where the effective pressure $P_{eff}(r)$ is defined as
\begin{equation}
P_{eff}(r) \equiv P(r)+P_{DE}(r).
\end{equation}
In this formulation, the charge dependent term is manifest, and the conservation equation reduces to the uncharged case when $Q=0$.


The Einstein field equations for the $G_{\phantom{1}t}^{t}$, $G_{\phantom{1}%
r}^{r}$ and $G_{\phantom{1}\theta _{i}}^{\theta _{i}}$ components then become
\begin{widetext}
\begin{eqnarray}
 \frac{(D-2)\lambda^{\prime }e^{-\lambda}}{2r}-\frac{%
(D-2)(D-3)(e^{-\lambda}-1)}{2r^2} &=& \kappa\left(\rho c^2+ \rho_{DE}c^2+\frac{1}{2}\frac{Q^2(r)}{r^{2(D-2)}}\right), 			 \label{00} \\
 \frac{(D-2)\nu^{\prime }e^{-\lambda}}{2r}+\frac{%
(D-2)(D-3)(e^{-\lambda}-1)}{2r^2} &=& \kappa\left(P+ P_{DE}-\frac{1}{2}\frac{Q^2(r)}{r^{2(D-2)}}\right),				 \label{rr} \\
 e^{-\lambda}\left( \frac{\nu^{\prime \prime }}{2}+\frac{%
\nu^{\prime 2}}{4}-\frac{\nu^{\prime }\lambda^{\prime }}{4}+\frac{%
(D-2)(\nu^{\prime }-\lambda^{\prime })}{4r} \right)&+& \frac{%
(D-3)(D-4)(e^{-\lambda}-1)}{2r^2}
= \kappa\left(P+ P_{DE}+\frac{1}{2}\frac{Q^2(r)}{r^{2(D-2)}}\right),   \label{ii}
\end{eqnarray}
\end{widetext}
respectively.

\subsection{The TOV equation for a charged sphere in $D-$dimensional space-times } \label{Sec.2.1}

 Equation~(\ref{00}) can be integrated immediately to obtain the
`accumulated' mass $M(r)$ inside radius the $r$, which is a function of the cosmological constant, $\Lambda_D$,
and of the charge integral,
\begin{eqnarray}\label{eq39}
\frac{2\kappa c^{2}}{(D-2)\Omega _{D-2}}M(r) &=&r^{D-3}(1-e^{-\lambda (r)})-
\notag \\
&&\frac{2\Lambda _{D}}{(D-1)(D-2)}r^{D-1}- \nonumber\\
&&\frac{\kappa }{(D-2)}\int_{0}^{r}\frac{Q^{2}(r^{\prime })}{r^{\prime D-2}}
dr^{\prime }.
\end{eqnarray}

The charge integral in Eq.~(\ref{eq39}) can be transformed by integration by parts, giving
\begin{eqnarray}
\frac{1}{r^{D-3}}\int_{0}^{r}\frac{Q^{2}(r^{\prime })}{r^{\prime D-2}}
dr^{\prime } &=&\frac{-1}{(D-3)r^{D-3}}\Bigg(\frac{Q^{2}(r)}{r^{D-3}}-
\notag \\
&&\int_{0}^{r}\frac{2Q(r^{\prime })}{r^{\prime D-3}}\frac{dQ(r^{\prime })}{
dr^{\prime }}dr^{\prime }\Bigg).  \label{IBP}
\end{eqnarray}
In the integral no surface term at infinity appears and the result is not valid for all space, but only for $r<R$, where $R$ is the radius of the sphere.
The second term on the right-hand side of Eq.~(\ref{IBP}) is the electromagnetic mass contribution, to be included in the total mass of the sphere.
Thus, $e^{-\lambda (r)}$ becomes
\begin{widetext}
\begin{eqnarray}
e^{-\lambda(r)} &=& 1 - \frac{2\kappa c^{2}}{(D-2)\Omega_{D-2}}\frac{M(r)}{r^{D-3}}+\frac{\kappa}{(D-2)(D-3)} \frac{Q^2(r)}{r^{2(D-3)}}   \nonumber  \\
			&\phantom{1}&   -\frac{2\Lambda_{D}}{(D-1)(D-2)}r^2  -\frac{2\kappa c^{2}}{(D-2)\Omega_{D-2}}\frac{\Omega_{D-2}}{(D-3)c^2r^{D-3}}\int_{0}^{r} \frac{Q(r')}{r'^{D-3}}\frac{dQ(r')}{dr'}dr'.
\end{eqnarray}
\end{widetext}

We now define the {\it total mass} $M_T(r)$ as the sum of the {\it matter mass} $M(r)$ and of the {\it electromagnetic mass} generated by the charge $Q(r)$ as
\begin{equation}
M_T(r) = M(r) + \frac{\Omega_{D-2}}{(D-3)c^2}\int_{0}^{r} \frac{Q(r^{\prime
})}{r^{\prime D-3}}\frac{dQ(r^{\prime })}{dr^{\prime }}dr^{\prime }. \label{totmass}
\end{equation}
The total mass-energy density inside the fluid sphere, including the electromagnetic contribution $\rho_{q}(r)$, can then be defined as
\begin{eqnarray}
\rho_{T}(r) &=& \frac{1}{\Omega_{D-2}r^{D-2}}\frac{d}{dr}\Big( M(r) +  \notag \\
&&\frac{\Omega_{D-2}}{(D-3)c^2}\int_{0}^{r} \frac{Q(r^{\prime })}{r^{\prime
D-3}}\frac{dQ(r^{\prime })}{dr^{\prime }}dr^{\prime }\Big)  \notag \\
&=& \rho(r) + \frac{1}{(D-3)c^2}\frac{Q(r)}{r^{2D-5}}\partial_r Q(r)  \notag
\\
&\equiv& \rho(r) + \rho_{q}(r),
\end{eqnarray}
where $\rho_{q}(r)$ has been defined implicitly. 

Using the definition of $M_T(r)$, $e^{-\lambda(r)}$ can then be written in the simpler form
\begin{eqnarray}
e^{-\lambda(r)} &=& 1-\frac{2\kappa c^{2}}{(D-2)\Omega_{D-2}}\frac{M_{T}(r)}{%
r^{D-3}}+  \notag \\
&&\frac{2\kappa}{(D-2)(D-3)}r^2\mathcal{U}(r) -\frac{2\Lambda_{D}}{(D-1)(D-2)}%
r^2,  \notag \\
\end{eqnarray}
where
\begin{equation}
\mathcal{U}(r) \equiv \frac{Q^2(r)}{2r^{2(D-2)}}.
\end{equation}
By substituting the above expression into Eq. (\ref{rr}), we obtain
\begin{widetext}
\begin{eqnarray}
\frac{e^{-\lambda(r)}}{r}\nu'(r) &=& \frac{2\kappa}{(D-2)}\left[ P_{eff}(r)-\mathcal{U}(r) \right]
\phantom{1}   +(D-3)\Bigg[ \frac{2\kappa c^{2}}{(D-2)\Omega_{D-2}}\frac{M_{T}(r)}{r^{D-1}}-\frac{2\kappa}{(D-2)(D-3)}\mathcal{U}(r)  \nonumber\\ &&+\frac{2\Lambda_{D}}{(D-1)(D-2)} \Bigg].    \label{TOV}
\end{eqnarray}
\end{widetext}
Thus, with the use of Eqs.~(\ref{conservation}) and (\ref{TOV}) we obtain the TOV equation, describing the structure of a charged sphere in $D$ dimensions, which takes the form
\begin{widetext}
\begin{eqnarray}  \label{TOVeqn}
\frac{dP_{eff}}{dr}&=&-\frac{(P_{eff}+\rho c^{2}+\rho_{DE}c^{2})}{(D-2)r^{D-2}y^{2}}
\left[(\kappa P_{eff}-\frac{2\Lambda_{D}}{D-1})r^{D-1}+(D-3)\frac{\kappa c^{2}M_{T}(r)}{\Omega_{D-2}}-\frac{\kappa Q^{2}}{r^{D-3}} \right]
+\frac{QQ'}{r^{2(D-2)}},
\end{eqnarray}
\end{widetext}
where we have introduced the Buchdahl variables,  defined as
\begin{eqnarray}
&&y^2 \equiv e^{-\lambda(r)} = 1-2w(r)r^2 + \frac{2\kappa}{(D-2)(D-3)}%
\mathcal{U}(r)r^2 ,  \notag \\
&&\zeta(r) = e^{\nu(r)/2}, \quad \quad x = r^2,  \notag \\
&&w(r) = \frac{\kappa c^{2}}{(D-2)\Omega_{D-2}}\frac{M_{T}(r)}{r^{D-1}}+
\frac{\Lambda_{D}}{(D-1)(D-2)}.  \notag
\end{eqnarray}
We note that the right-hand side of the TOV equation contains an extra term $QQ^{\prime }/r^{2(D-2)}$, vis-{\'a}-vis the uncharged case \cite{Burikham:2015nma},
which represents the charge contribution to the hydrostatic equilibrium.

\section{Mass limits in $D$ dimensions for charged spherically symmetric objects in the presence of dark energy} \label{Sec.3}

In terms of Buchdahl variables, Eqs.~(\ref{TOV}) and (\ref{conservation}) can be written
as
\begin{eqnarray}
\frac{\zeta^{\prime }(x)}{\zeta} &=& \frac{\nu^{\prime }(x)}{2}, \\
&=& \frac{1}{y^2}\left[ \frac{4\pi G_D}{(D-2)c^4}P_{eff}(x) + \frac{D-3}{2}w(x) -\frac{\kappa \mathcal{U}}{D-2} \right], \notag
\label{TOV2}
\end{eqnarray}
and
\bea
P^{\prime }_{eff}(x) &=& -\left[\rho(x)c^{2}+\rho_{DE}c^{2}+P_{eff}(x)\right]%
\frac{\zeta^{\prime }(x)}{\zeta} \nonumber\\
&+&\frac{Q(x)}{x^{D-2}}Q^{\prime }(x),
\label{conservation2}
\eea
respectively, where $(\ldots)^{\prime }(x)$ denote the differentiation with respect to $x$.
More elegantly, Eq.~(\ref{conservation2}) can be formulated in another way as
\begin{equation}
(P_{eff}\zeta)^{\prime }(x) = -(\rho+\rho_{DE})c^{2}\zeta^{\prime }(x) +
\frac{Q(x)}{x^{D-2}}Q^{\prime }(x)\zeta(x).  \label{conservation3}
\end{equation}
This relation will be important later, in the derivation of the Buchdahl
inequality. Furthermore, there exists an additional relation between $\rho_{T},
\rho_{DE}$ and the function $w(x) = w(r^2)$, defined as
\be
w(x) = \frac{\kappa}{2(D-2)}\frac{1}{x^{\frac{D-1}{2}}} \int_0^x
(\rho_{T}c^2+\rho_{DE}c^2)x^{\prime \frac{D-3}{2}}dx^{\prime },
\ee
namely
\be
(x^{\frac{D-1}{2}}w(x))^{\prime }= \frac{\kappa}{2(D-2)}%
(\rho_Ec^2+\rho_{DE}c^2)x^{\frac{D-3}{2}}.  \label{diffw}
\ee

From Eq.~(\ref{TOV2}), we then obtain
\begin{equation}\label{eq53}
y^2\zeta^{\prime }= \frac{\kappa}{2(D-2)} P_{eff}\zeta + \frac{D-3}{2}w\zeta - \frac{\kappa ~\mathcal{U}\zeta}{D-2}.
\end{equation}
After differentiating Eq.~(\ref{eq53}) with respect to $x$ (and with the use of Eq.~(\ref{conservation3})), we find
\begin{eqnarray}
\left(y^2\zeta^{\prime }\right)^{\prime }&=& -\frac{\kappa}{2(D-2)}(\rho+\rho_{DE})%
\zeta^{\prime } -\frac{\kappa}{D-2}(\mathcal{U}\zeta)' \nonumber\\
&&+\frac{\kappa}{2(D-2)}\frac{Q}{x^{D-2}}Q^{\prime }\zeta + \frac{D-3}{2}%
(w\zeta)^{\prime }.
\eea
Since $\rho_{T} = \rho + \rho_{q}$, we next obtain
\bea
\left(y^2\zeta^{\prime }\right)^{\prime }&=& -\frac{\kappa}{2(D-2)}(\rho_{T}+\rho_{DE})\zeta^{\prime } -\frac{\kappa}{D-2}(\mathcal{U}\zeta)'  \nonumber \\
&&+\frac{\kappa}{2(D-2)}\rho_{q}\zeta^{\prime }+ \frac{\kappa}{2(D-2)}\frac{Q%
}{x^{D-2}}Q^{\prime }\zeta  \nonumber\\
&&+\frac{D-3}{2}(w\zeta)^{\prime },
\eea
and, with the use of  Eq.~(\ref{diffw}), we arrive at the result
\bea
\left(y^2\zeta^{\prime }\right)^{\prime }&=& -\frac{1}{x^{\frac{D-3}{2}}}\left(x^{\frac{D-1}{2}}w\right)^{\prime }\zeta^{\prime
}+ \frac{\kappa}{2(D-2)}\rho_{q}\zeta^{\prime }  \notag \\
&&+\frac{\kappa}{2(D-2)}\frac{Q}{x^{D-2}}Q^{\prime }\zeta + \frac{D-3}{2}%
(w\zeta)^{\prime }   \notag \\
&& -\frac{\kappa}{D-2}(\mathcal{U}\zeta)'.  \label{zeta1}
\end{eqnarray}
Furthermore, since $y(x)$ is defined in terms of Buchdahl variables as
\begin{equation}
y^2(x) = 1-2w(x)x+\frac{2\kappa}{(D-2)(D-3)}\mathcal{U}(x)x,  \label{zeta2}
\end{equation}
we have
\bea
2yy^{\prime }&=& -2w^{\prime }x-2w +\frac{2\kappa}{(D-2)(D-3)}\mathcal{U}%
^{\prime }x
\nonumber\\
&+ &\frac{2\kappa}{(D-2)(D-3)}\mathcal{U},  \label{zeta3}
\eea
where
\be\label{zeta4}
\mathcal{U}(x) = \frac{Q^2}{2x^{D-2}}, \rho_{q}(x) = \frac{2}{D-3}\left(
\frac{Q}{x^{D-3}}Q^{\prime }\right),
\ee
and
\be
\mathcal{U}^{\prime }(x) = \frac{Q}{x^{D-2}}Q^{\prime }-\frac{D-2}{x}%
\mathcal{U}.  \label{zeta5}
\ee
Using Eqs.~(\ref{zeta2})-(\ref{zeta5}) in Eq.~(%
\ref{zeta1}), we finally obtain
\begin{equation}
y\left(y\zeta^{\prime }\right)^{\prime }= \frac{D-3}{2}w^{\prime }\zeta + \frac{\kappa Q^{2}\zeta}{
2 x^{D-1}} - \frac{\kappa QQ'}{2(D-2)x^{D-2}}.  \label{buceqn}
\end{equation}
This is the generalized Buchdahl equation for spherically symmetric, charged compact objects in $D-$dimensional space-time.
As compared to the uncharged case \cite{Burikham:2015nma}, it contains an additional term on the right-hand side, which is the extra contribution due to the presence of the
electric charge.

\subsection{Buchdahl inequality in $D$ space-time dimensions}

In the following, we introduce four new variables $z$, $\gamma$, $\psi$ and $\eta $, defined as
\begin{equation}\label{eq62}
dz=\frac{1}{y(x)}dx \to z(r) = \int_0^r \frac{2r^{\prime }}{ \sqrt{ 1 -
\frac{\Theta(r^{\prime })}{r^{\prime D-3}}}} dr^{\prime }, \\
\end{equation}
\begin{equation}
\gamma(r) \equiv \frac{Q^{2}\zeta}{r^{2D-3}},  \label{gam}
\end{equation}
\begin{equation}
\psi = \zeta - \eta,
\end{equation}
where $\eta(r)$ is defined in
terms of the integral
\begin{equation}
\eta(r) = 2\kappa \int_0^r \left( \int_0^{r_1} \frac{\gamma(r_2)%
} {\sqrt{1-\frac{\Theta(r_2)}{r_2^{D-3}} } } dr_2 \right) \frac{r_1}{
\sqrt{1-\frac{\Theta(r_1)}{r_1^{D-3}} } } dr_1.
\end{equation}
The function  $\Theta(r)$, introduced in Eq.~(\ref{eq62}), is explicitly defined by
\begin{equation}
\Theta(r) \equiv \frac{2\kappa c^{2}M_{T}(r)}{(D-2)\Omega_{D-2}}-\frac{%
2\kappa~\mathcal{U}(r)r^{D-1}}{(D-2)(D-3)}+\frac{2\Lambda_{D} r^{D-1}}{(D-1)(D-2)%
},
\end{equation}
giving $y^{2}=1-\Theta(r)/r^{D-3}$.
In terms of the new variables introduced above, the Buchdahl equation Eq.~(\ref{buceqn}) can be written as
\begin{equation}
\frac{d^2\psi(z)}{dz^2} = \frac{D-3}{2}w^{\prime }(x)\zeta(x) - \frac{\kappa QQ'}{2(D-2)x^{D-2}}.  \label{buch}
\end{equation}

For a stable charged object, the assumption that the average total density
\begin{eqnarray}
\bar{\rho}_{T}&=&\frac{(D-1)M_{T}(r)}{\Omega_{D-2}r^{D-1}},
\end{eqnarray}
does not increase with $r$ implies that $M_{T}/r^{D-1}$ is a decreasing
function. Therefore, we assume that, for all $r^{\prime }< r $, $\Theta(r)$
satisfies the condition
\begin{equation}
\frac{\Theta(r^{\prime })}{r^{\prime D-3}} \geq \frac{\Theta(r)}{%
r^{D-3}}\left( \frac{r^{\prime }}{r} \right)^2,  \label{cond1}
\end{equation}
and also that
\begin{equation}
\gamma(r^{\prime }) \geq \gamma(r).  \label{cond2}
\end{equation}
From Eq.~(\ref{buch}), we can then obtain the generalized Buchdahl inequalities from the
condition that, for any physical density profile of the charged object, the following relations hold,
\begin{equation}
\frac{d}{dr}\frac{M_{T}(r)}{r^{D-1}} < 0 \quad \to \quad w^{\prime }(x)<0,
\end{equation}
together with the condition that $(Q^{2})'>0$ inside the object.
The inequality
\begin{equation}
\frac{d^2}{dz^2}\psi(z) < 0
\end{equation}
therefore holds, for all $r$ in the range $0 \leq r \leq R$, where $R$ is the radius of the fluid sphere, for any static charged object.

Using the mean value theorem, we get
\begin{equation}
\frac{d}{dz}\psi \leq \frac{\psi(z)-\psi(0)}{z},
\end{equation}
and, since $\psi(0) = \zeta(0)-\eta(0) = \zeta(0) > 0$, it follows that
\begin{equation}
\frac{d}{dz}\psi(z) \leq \frac{\psi(z)}{z} \quad \to \quad \frac{d}{dz}\zeta
- \frac{d}{dz}\eta \leq \frac{\zeta - \eta }{z}.
\end{equation}
The above relation can be written explicitly as
\begin{widetext}
\begin{eqnarray}
&&\frac{1}{2r}\sqrt{1-\frac{\Theta(r)}{r^{D-3}} }\frac{d\zeta}{dr} - \kappa \int_0^r \frac{\gamma(r')}{\sqrt{1-\frac{\Theta(r')}{r'^{D-3}} }} dr' \leq \nonumber \\
&& \frac{1}{ 2\int_0^r \frac{r'}{\sqrt{1-\frac{\Theta(r')}{r'^{D-3}} }} dr'}\left[ \zeta - 2\kappa \int_0^{r} \frac{r_1}{\sqrt{1-\frac{\Theta(r_1)}{r_1^{D-3}} }}  \left( \int_0^{r_1} \frac{\gamma(r_2)}{\sqrt{1-\frac{\Theta(r_2)}{r_2^{D-3}} }} dr_2 \right) dr_1 \right].    \label{buch2}
\end{eqnarray}
\end{widetext}
All the integrals in the inequality (\ref{buch2}) can be evaluated using the conditions (\ref%
{cond1}) and (\ref{cond2}). By using (\ref{cond1}), it follows that the denominator of right-hand side
of (\ref{buch2}) is bounded, such that
\begin{equation}
\left( \int_0^r \frac{r^{\prime }}{\sqrt{1-\frac{\Theta(r^{\prime })}{%
r^{\prime D-3}} }} dr^{\prime }\right)^{-1} \leq \frac{\Theta(r)}{%
r^{D-1}}\left(1-\sqrt{1-\frac{\Theta(r)}{r^{D-3}} } \right)^{-1}.
\label{int1}
\end{equation}
The second term on the left-hand side of (\ref{buch2}) also has an upper bound,
\begin{eqnarray}
\phantom{1} \int_0^r \frac{\gamma(r^{\prime })}{\sqrt{1-\frac{\Theta%
(r^{\prime })}{r^{\prime D-3}} }} dr^{\prime }&\leq& \gamma_{0}\int_0^r
\frac{1}{y} dr^{\prime }= \frac{\gamma_{0}r}{y},
\end{eqnarray}
where $\gamma_{0}\equiv \gamma(r=0)$ is the central value of $\gamma $. The term involving $%
\gamma(r)$ also has a lower bound,
\begin{eqnarray}  \label{int2}
\phantom{1} \int_0^r \frac{\gamma(r^{\prime })}{\sqrt{1-\frac{\Theta%
(r^{\prime })}{r^{\prime D-3}} }} dr^{\prime }&\geq& \gamma(r) \int_0^r
\left( 1- \frac{\Theta(r)}{r^{D-3}}\left( \frac{r^{\prime }}{r}
\right)^2 \right)^{-\frac{1}{2}} dr^{\prime }  \notag \\
&=& \gamma(r) \left( \frac{\Theta(r)}{r^{D-1}}\right)^{-\frac{1}{2}%
}\arcsin{\left(\sqrt{\frac{\Theta(r)}{r^{D-3}}} \right)}.  \notag \\
\end{eqnarray}
Consequently, the term in the numerator on the right-hand side of (\ref{buch2}) is
bounded from below by
\begin{widetext}
\begin{eqnarray}
&\phantom{1}&  \int_0^{r} \frac{r_1}{\sqrt{1-\frac{\Theta(r_1)}{r_1^{D-3}} }}  \left( \int_0^{r_1} \frac{\gamma(r_2)}{\sqrt{1-\frac{\Theta(r_2)}{r_2^{D-3}} }} dr_2 \right) dr_1
\geq \int_0^r  r_1\left( 1- \frac{\Theta(r_1)}{r_1^{D-3}} \right)^{-\frac{1}{2}}  \left( \frac{\gamma(r_1)}{\left( \frac{\Theta(r_1)}{r_1^{D-1}} \right) ^{\frac{1}{2}}} \arcsin{ \left({\sqrt{\frac{\Theta(r_1)}{r_1^{D-3}} } } \right)} \right)dr_1,   \nonumber \\
&\geq& \frac{\gamma(r)}{\left( \frac{\Theta(r)}{r^{D-1}} \right)^{\frac{1}{2}}} \int_0^r  r_1 \left( 1- \frac{\Theta(r)}{r^{D-1}}r_1^2 \right)^{-\frac{1}{2}} \arcsin{ \left({\sqrt{\frac{\Theta(r)}{r^{D-1}} }r_1 } \right)}dr_1
=  \frac{\gamma(r)}{\left( \frac{\Theta(r)}{r^{D-1}} \right)^{\frac{3}{2}}} \int_0^s    \frac{ s'}{(1-  s'^2)^{\frac{1}{2}}} \arcsin{ s'}d s'  \nonumber \\
&=&  \frac{\gamma(r)}{\left( \frac{\Theta(r)}{r^{D-1}} \right)^{\frac{3}{2}}} \Bigg{[}  \left(\frac{\Theta}{r^{D-1}}\right)^{\frac{1}{2}}r - \sqrt{1-\left(\frac{\Theta}{r^{D-1}}\right)r^2}\arcsin\left( \left(\frac{\Theta}{r^{D-1}}\right)^{\frac{1}{2}}r\right) \Bigg{]},    \label{int3}
\end{eqnarray}
\end{widetext}
where we have denoted
\begin{equation}
s^{\prime }\equiv \left(\frac{\Theta}{r^{D-1}}\right)^{\frac{1}{2}}r_1,
s \equiv \left(\frac{\Theta}{r^{D-1}}\right)^{\frac{1}{2}}r.
\end{equation}

Plugging integrals (\ref{int1}), (\ref{int2}) and (\ref{int3}) into (\ref%
{buch2}) and using the relation $y^{2}=1-\Theta(r)/r^{D-3}$, we next obtain
\begin{widetext}
\begin{eqnarray}
\frac{\nu'(r)}{2r}\zeta(r)y(r) &\leq& \frac{1+y}{r^{2}}\left[ \zeta(r) -2\kappa\gamma(r)r^{3}\left( \frac{1}{1-y^{2}}-\frac{y\arcsin{(\sqrt{1-y^{2}})}}{(1-y^{2})^{3/2}}\right) \right]
+ 2\kappa\frac{\gamma_{0}r}{y}.         \label{buch3}
\end{eqnarray}
\end{widetext}
At the surface of the compact object, $r=R$, and thus, using Eq.~(\ref{TOV2}) together with the value of the dark energy at the surface $%
P_{eff}(r=R)=w\rho_{DE}c^{2}\equiv P_{e}$, we find
\begin{widetext}
\begin{eqnarray}
\left( \frac{\kappa}{D-2}P_{e}+(D-3)w(R) -\frac{2\kappa \mathcal{U}}{(D-2)}\right)R^{2}
&\leq &y^{2}+y
+2\kappa R^{3}\left[ \frac{\gamma_{0}}{y}+\frac{\gamma(R)}{1-y}%
\left( \frac{y\arcsin{(\sqrt{1-y^{2}})}}{\sqrt{1-y^{2}}}-1\right)\right]\nonumber\\
&&\leq y^{2}+y+2\kappa R^{3}\frac{\gamma_{0}}{y},  \label{buch4}
\end{eqnarray}
\end{widetext}
where we have used the fact that $\arcsin{(\sqrt{1-y^{2}})}\leq \sqrt{1-y^{2}}/y$, and have assumed that $\gamma
\geq 0$ for $0\leq r\leq R$.  In the $\gamma_{0}$ term, we also assume that $\zeta(R) \geq y(R)$ as a result of the energy condition $\rho c^{2} +P\geq 0$, which allows us to replace $1/\zeta$ with $1/y$.

For $\gamma_{0}=0$, the inequality (\ref{buch4}) properly reduces to the
uncharged case discussed in \cite{Burikham:2015nma}. Even when
$y$ is bounded between 0 and 1, the last term on the left-hand side is unbounded, since
it is proportional to $1/y$. The metric becomes the black hole metric for $%
y=0$, but we expect the Buchdahl limit to set in before this point, at the upper bound
$\Theta/R^{D-3}=1-y^{2}$.

\subsection{Dimensionless form of the mass bounds}

We now introduce a new set of dimensionless variables $\{\Gamma, u,b,\Delta\}$, defined as
\begin{equation}
\Gamma \equiv 2\kappa R^{3}\gamma_{0}\left(\frac{1}{y}%
\right)_{upper},  \label{Gam0}
\end{equation}
and
\begin{eqnarray}
&&u\equiv \frac{\kappa c^{2}M_{T}}{(D-2)\Omega_{D-2}R^{D-3}}, ~b\equiv \frac{%
\Lambda_{D} R^{2}}{D-2},  \notag \\
&&\Delta \equiv \frac{2\kappa R^{2}\mathcal{U}}{(D-2)(D-3)}=\frac{\kappa
Q^{2}}{(D-2)(D-3)R^{2(D-3)}},  \nonumber\\  \label{dlpar}
\end{eqnarray}
respectively. In terms of these dimensionless quantities, the inequality (\ref{buch4}) can be expressed in a simple form  as
\begin{eqnarray}\label{fb}
(D-1)u+(1+w)b-(D-2)\Delta \leq 1+y+\Gamma,
\end{eqnarray}
where $y=\sqrt{1-(2u+2b/(D-1)-\Delta)}$. The inequality (\ref{fb}) implies the bounds
\begin{eqnarray}
u_{-}\leq u \leq u_{+},  \label{ubounds}
\end{eqnarray}
where
\begin{eqnarray}  \label{mbs}
u_{\pm}&=& -\frac{B}{2A}\left( 1\pm \sqrt{1-\frac{4AC}{B^{2}}} \right), \ A = (D-1)^{2},
\label{uform} \\
B&=&2+2(D-1)[b(1+w)-(1+(D-2)\Delta+\Gamma)],  \notag \\
C &=& -(1+\Delta)+\frac{2b}{D-1} \nonumber\\
&+&\Big[b(1+w)-(1+(D-2)\Delta+\Gamma)\Big]^{2}.  \notag
\end{eqnarray}

As a cross-check of our main result, we compare these bounds with the known result for uncharged
objects obtained in \cite{Burikham:2015nma}. When $Q=0$, we have $\Delta = 0$, $ \Gamma =0$,
and
\begin{equation}
M \equiv M_{T}\vert_{Q=0} \to u = \frac{8\pi G_DM}{(D-2)\Omega_{D-2}R^{D-3}}.
\end{equation}
In this case,
\begin{widetext}
\begin{equation}
u_{\pm }=\frac{D-2}{(D-1)^{2}}\left[ 1-(1+w)\frac{D-1}{(D-2)^{2}}\Lambda
_{D}R^{2}\right] \pm \frac{D-2}{(D-1)^{2}}\sqrt{1+2w\frac{D-1}{(D-2)^{3}}
\Lambda _{D}R^{2}},  \label{olims}
\end{equation}
\end{widetext}
in complete agreement with the previous result \cite{Burikham:2015nma}.

In order to obtain the influence of the charge on the mass bounds, we need to
evaluate the value of $\Gamma$, which contains the unknown parameter $\gamma_{0}$
and the upper bound $u_{+}$~(through $(1/y)_{upper}$ in (\ref{Gam0})).
Since $\zeta(r=0) =1$,
an estimate value of $\gamma_{0}$ can be obtained from Eq.~(\ref{gam}) as
\begin{eqnarray}
\gamma_{0}&\approx& \frac{e^{2}}{\lambda_{e}^{2D-3}},
\end{eqnarray}
where $\lambda_{e}=\hbar/m_{e}c$ is the Compton wavelength of the electron.
This is equivalent to the statement that a charge $e$  cannot be compressed
to within a radius smaller than $\lambda_{e}$. Consequently, we can
approximate
\begin{eqnarray}
\Gamma&\approx& 2\kappa R^{3}\frac{e^{2}}{\lambda_{e}^{2D-3}}%
\left[ 1-\left(2u_{+}+\frac{2b}{D-1}-\Delta\right)\right]^{-1/2}. \notag \\ \label{Gam}
\end{eqnarray}
The dimensionless quantity $\Gamma \sim R^{3}$ can have values as small as $10^{-87}$%
~(for $M_{T}\simeq 2\times 10^3$ g, $R=1.5R_{S}$, where $R_{S}=2GM/c^{2}$ is the
corresponding Schwarzschild radius) to as large as $1600$~(for $M_{T}\simeq
M_{\odot}=2\times 10^{33}$ g, $R=1.5R_{S}$) when $D=4$ and for very small $\Lambda$. Thus,
we cannot take it to be a generically small quantity.

On the other hand, the cosmological constant in four space-time dimensions is extremely small in the
real physical world. The quantity $b$ is thus very small, since $R$ is
typically much less than, or at most comparable to the size of the universe, $R \lesssim R_U \approx R_W = \sqrt{3/\Lambda}$. Therefore, it is interesting to
explore the mass bounds for charged spherical objects when $\Delta, \Gamma
\gg b$. Another important limit is when the dark energy becomes a
cosmological constant with $w=-1$. We will explore mass limits in these
situations in the following Section.

\section{Mass limits for $\Lambda R^{2}\ll1$ cases} \label{Sec.4}

In this Section, we assume the dark energy density to be very small, i.e. $%
\Lambda R^{2}\ll 1$. We first consider the case in which dark energy corresponds to a cosmological constant with $w=-1$.
The general case, with arbitrary $w$, is then investigated.

\subsection{Cosmological constant dark energy}  \label{Sec.4.1}

If dark energy corresponds to a cosmological constant and $b$ is very small, we
can set $w=-1$ in Eq.~(\ref{mbs}) so that
\begin{widetext}
\begin{eqnarray}
\sqrt{1-\frac{4AC}{B^{2}}}&\simeq& \sqrt{1-\frac{(D-1)^{2}[(1+(D-2)\Delta+\Gamma)^{2}-(1+\Delta)]}{[(D-1)(1+(D-2)\Delta+\Gamma)-1]^{2}}}
\nonumber\\
&&\times\left[ 1-\frac{b}{(D-1)}\left(1+\Delta+\frac{1}{(D-1)^{2}}-\frac{2}{D-1}(1+(D-2)\Delta+\Gamma)\right)^{-1}\right],
\end{eqnarray}
\end{widetext}
while
\begin{eqnarray}
-\frac{B}{2A}&=& \frac{(D-1)(1+(D-2)\Delta+\Gamma)-1}{(D-1)^{2}}>0.
\end{eqnarray}
The upper~(lower) mass limit is then given by $u_{+(-)}$ respectively. For $%
B<0, C>0$, $u_{-}$ is positive and the lower limit exists. It is
straightforward to see that $B$ is always negative for $D\geq 2$ and $C$ is
always positive for $\Lambda_D \geq 0$, since $\Delta, \Gamma \geq 0$.

\subsection{Generic dark energy}  \label{Sec.4.2}

A more general situation is when we allow a small $w$-dependent term to
exist for $\omega \neq -1$. In this case, we may approximate the quantity $\sqrt{1-4AC/B^2}$ as
\begin{widetext}
\begin{eqnarray}\label{mlimb1}
\sqrt{1-\frac{4AC}{B^{2}}}&\simeq& \sqrt{1-\frac{4AC_{0}}{B_{0}^{2}}}  \\
&-&\frac{b(D-1)^{2}}{ \sqrt{1-\frac{4AC_{0}}{B_{0}^{2}}}}\left\{ \frac{\frac{1}{D-1}-(1+w)(1+(D-2)\Delta +\Gamma)}{[(D-1)(1+(D-2)\Delta+\Gamma)-1]^{2}}
- \frac{(D-1)(1+w)[(1+(D-2)\Delta+\Gamma)^{2}-(1+\Delta)]}{[(D-1)(1+(D-2)\Delta+\Gamma)-1]^{3}} \right\},    \notag
\end{eqnarray}
\end{widetext}
where
\begin{eqnarray} \label{mlimb2}
B_{0} &\equiv& B(b=0)=2-2(D-1)(1+(D-2)\Delta+\Gamma),
\nonumber\\
C_{0} &\equiv& C(b=0)=(1+(D-2)\Delta+\Gamma)^{2}-(1+\Delta),
\nonumber\\
-\frac{B}{2A}&=&-\frac{(D-1)[b(1+w)-(1+(D-2)\Delta+\Gamma)]+1}{(D-1)^{2}}. \notag
\end{eqnarray}
Since $C_{0}>0$, the second term on the right-hand side of Eq. (\ref{mlimb1}) will determine
whether the lower bound $u_{-}$ exists. Thus, for $\Lambda >(<) \ 0$, a nontrivial positive lower bound $u_{-}$ will exist if
\begin{eqnarray}
w & <(>)& \frac{1-(D-1)(1+(D-2)\Delta+\Gamma)-\frac{2AC_{0}}{B_{0}}}{%
(D-1)(1+(D-2)\Delta+\Gamma)+\frac{2AC_{0}}{B_{0}}}. \notag \\
\end{eqnarray}

\subsection{Charged sphere with no dark energy}  \label{Sec.4.3}

We may also set $b=0$ and simply consider the mass limits for a charged sphere in
asymptotically flat space. From Eq.~(\ref{mbs}), it follows that
\begin{eqnarray}
\sqrt{1-\frac{4AC}{B^{2}}}&=& \sqrt{1-\frac{4AC_{0}}{B_{0}^{2}}}<1,
\end{eqnarray}
and
\begin{eqnarray}
-\frac{B}{2A}&=&\frac{(D-1)[(1+(D-2)\Delta+\Gamma)]-1}{(D-1)^{2}}.  \label{mlimq}
\end{eqnarray}
Hence, for $D\geq2$, both the upper and lower limits always exist according to Eq. (\ref{mlimq}).

Due to the electrostatic repulsion between charged fluid elements, the minimum mass/radius power ratio is required in order for gravitational attraction to counteract the repulsive force, which enables the object to maintain a static configuration. The maximum mass/radius power ratio denotes the limit before gravitational collapse. For a charged object, we expect the maximum mass/radius power to be greater than that for an uncharged object, i.e. the Buchdahl limit Eq. (\ref{Buch}), due to the repulsive effect of the charge density.

\subsection{Small objects with $\Delta, \Gamma \ll 1$}  \label{Sec.4.4}

As long as the size $R$ of the object is sub-astronomical (i.e. small compared to the size of a typical star), the numerical value of $\Gamma$ in $(3+1)$ dimensions is generically much smaller than unity. We now  consider the condition $\Gamma \ll 1$, together with $\Delta \ll 1$, in \emph{arbitrary} dimensions, and derive explicit expressions for the mass bounds in this limit.

For $\Delta, \Gamma \ll 1$, we can approximate $\sqrt{1-4AC/B^2}$ as
\begin{eqnarray}
\sqrt{1-\frac{4AC}{B^{2}}}&\simeq& 1-\frac{(D-1)^{2}}{(D-2)^{2}} \\
&\times& \left[ \Gamma+\frac{(2D-5)\Delta}{2}-b\left(\frac{D-2}{D-1}+w\right) \right], \nonumber
\end{eqnarray}
and the mass limits $u_{\pm}$ become
\begin{widetext}
\begin{eqnarray}
u_{+}&\simeq&\frac{2(D-2)}{(D-1)^{2}}+\frac{1}{(D-1)(D-2)}\left( (2D^{2}-9D+11)\frac{\Delta}{2}+(D-3)\Gamma +[2+3w-D(1+w)]b \right), \\
u_{-}&\simeq&-\frac{1+(D-1)[b(1+w)-(1+(D-2)\Delta+\Gamma)]}{(D-2)^{2}}\left[ \Gamma+\frac{(2D-5)\Delta}{2}-b\left(\frac{D-2}{D-1}+w \right) \right].
\end{eqnarray}
\end{widetext}
The {\it minimum} mass/radius power can thus be approximated as
\begin{eqnarray}
\left(\frac{M_{T}}{R^{D-3}}\right)_{\rm min}&=&\frac{\Gamma\Omega_{D-2}}{\kappa c^{2}}+\frac{(2D-5)Q_{\rm tot}^{2}/\Omega_{D-2}}{2(D-2)(D-3)R^{2(D-3)}c^{2}}  \nonumber \\
&&-\frac{\Lambda_{D} R^{2}\Omega_{D-2}}{\kappa c^{2}}\left(\frac{w}{D-2}+\frac{1}{D-1}\right), \nonumber \\  \label{mrmin}
\end{eqnarray}
where we define the total charge contribution $Q_{\rm tot}\equiv \Omega_{D-2}Q$. Note again that the total mass $M_{T}$ is the sum of usual matter mass and the electromagnetic mass given by Eq.~(\ref{totmass}).
For zero dark energy, $b=0$, and for very small $\Gamma$ the minimum mass-power radius simplifies to
\begin{eqnarray}
\left(\frac{M_{T}}{R^{D-3}}\right)_{\rm min} &\approx& \frac{(2D-5)Q_{\rm tot}^{2}/\Omega_{D-2}}{2(D-2)(D-3)R^{2(D-3)}c^{2}}.
\end{eqnarray}
The minimum mass/radius power ratio can also be cast in the form of a minimum average density for a static spherical object, i.e
\begin{eqnarray}
\bar{\rho}_{\rm min}&=&\frac{\Gamma(D-1)}{\kappa c^{2}R^{2}}+\frac{Q^{2}(2D-5)(D-1)}{2(D-2)(D-3)R^{2(D-2)}c^{2}} \nonumber \\
&&-\frac{\Lambda_{D}}{\kappa c^{2}}\left( 1+w\frac{D-1}{D-2}\right).
\end{eqnarray}
Remarkably, with the contribution from the electric charge density, the lower bound~(i.e. nontrivial positive values of $u_{-}$) exists, for a wide range of values of $\Lambda_{D}$, in both the asymptotically de Sitter and anti-de Sitter cases. For positive~(negative) $\Lambda_D$, as long as $w$ is not much less~(more) negative than $-(D-2)/(D-1)$, the lower bound always exists.

On the other hand, the {\it maximum} mass/radius power ratio is
\begin{widetext}
\begin{eqnarray}
\left(\frac{M_{T}}{R^{D-3}}\right)_{\rm max}&=&\frac{2(D-2)^{2}\Omega_{D-2}}{(D-1)^{2}\kappa c^{2}}+\frac{\Gamma\Omega_{D-2}}{\kappa c^{2}}\left( \frac{D-3}{D-1}\right)+\frac{(2D^{2}-9D+11)Q_{\rm tot}^{2}/\Omega_{D-2}}{2(D-1)(D-2)(D-3)R^{2(D-3)}c^{2}} \label{mrmax} \\&&-\frac{\Lambda_{D} R^{2}\Omega_{D-2}}{\kappa c^{2}}\frac{[2+3w-D(1+w)]}{(D-1)(D-2)}. \nonumber
\end{eqnarray}
\end{widetext}
Thus, the maximum mass/radius power becomes larger in the presence of the electric charges, but the presence of dark energy could either enhance or weaken the upper bound, depending on the sign of $w$. For zero dark energy and very small $\Gamma$, the maximum mass/radius power limit reduces to
\begin{eqnarray}
\left(\frac{M_{T}}{R^{D-3}}\right)_{\rm max} &\approx& \frac{2(D-2)^{2}\Omega_{D-2}}{(D-1)^{2}\kappa c^{2}} \\
&&+\frac{(2D^{2}-9D+11)Q_{\rm tot}^{2}/\Omega_{D-2}}{2(D-1)(D-2)(D-3)R^{2(D-3)}c^{2}}. \nonumber
\end{eqnarray}

For convenience, we now present some results in $(3+1)$ dimensions with positive $\Lambda$ and $w=-1$.
\begin{eqnarray}
\left(\frac{M_{T}}{R}\right)_{\rm min}&=&\frac{3}{4}\left( \frac{4\pi Q^{2}}{R^{2}c^{2}}\right)+\frac{\Lambda R^{2}c^{2}}{12G}, \\
\left(\frac{M_{T}}{R}\right)_{\rm max}&=&\frac{4}{9}\frac{c^{2}}{G}+\frac{7}{12}\left( \frac{4\pi Q^{2}}{R^{2}c^{2}}\right)+\frac{\Lambda R^{2}c^{2}}{12G},
\end{eqnarray}
where we have assumed small contributions from the dimensionless quantity $\Gamma$. At this point, in order to facilitate the comparison between the previous four-dimensional results obtained in \cite{Boehmer:2007gq}, and the results of the present paper, we would like to mention that due to the different choice of units and scaling of physical parameters the results of \cite{Boehmer:2007gq} {\it can be re-obtained once the substitutions $4\pi Q^2/c^2\rightarrow Q^2$ and $\Lambda c^2\rightarrow 8\pi GB$ are performed in all the equations of the present paper}.

We end this Section with a discussion of the holographic interpretation of the minimum and maximum mass/radius power ratio, focusing on the scenario in which the object is embedded in
an asymptotically anti-de Sitter (AdS) space-time, with $\Lambda_{D}<0$. In this case, the maximum mass bound for a given radius of a charged object corresponds to the Hawking temperature, $T_{H}$, of a charged (i.e. Reissner-Nordstr{\" o}m) black hole~(RNBH), with the same mass.  At at this radius, any object with larger mass than the maximum mass will inevitably collapse to form a black hole.  For a large RNBH with positive heat capacity, $T_{H}$ is an increasing function with the black hole mass and can be determined once the mass is known~(see, for example, ~\cite{Burikham:2014gwa} for explicit formulae). From the viewpoint of holographic duality, the Hawking temperature can be identified with the temperature of the dual gauge plasma in the deconfined phase~(for example, the quark-gluon plasma in QCD). Specifically, it can be interpreted as the maximum temperature of the dual gauge matter in the confined phase \emph{before} it undergoes an inevitable phase transition into a deconfined phase. Or, in other words, as the confinement/deconfinement phase transition temperature.

Generically, a {\it static} configuration in the bulk gravity picture of the background AdS space is dual to a {\it thermal} phase in the boundary gauge picture.  To give a few examples: An empty bulk AdS space is dual to the confined phase of gauge matter on the boundary.  A black hole in the bulk is dual to the thermal phase of the gauge matter on the boundary, in which the Hawking temperature of the black hole is identified with the temperature of the thermal gauge phase. The mass of a static AdS star made of fermions is dual to the conformal dimension of the multitrace operator in the dual Conformal Field Theory~(CFT)~\cite{Arsiwalla:2010bt, Burikham:2012kn}.  In~\cite{Burikham:2014ova}, it was shown that the mass of the bulk AdS star is linearly proportional to the number density of particles on the boundary when the mass is large.  This provides a holographic correspondence between the bulk mass in the gravity picture and the particle density on the boundary in the gauge field picture.

With this in mind, we can interpret the minimum mass at a given radius, derived in this Section, to be the dual of the minimum density of the gauge matter living on the boundary space.  If the density of the gauge matter is too low, it will evaporate into a `hadron' gas.  This gauge picture corresponds to the gravity picture in which the mass of the spherical object scatters into the entire AdS space, since it is lower than the minimum mass required for stability at a given radius. Thus, the minimum mass/radius power ratio gives the critical density of the dual gauge `nucleus', under which it will evaporate into the `hadron' gas phase.

\subsection{Bounds on the static universe}  \label{Sec.5}

We can apply the condition (\ref{ubounds}) to the entire universe by setting $R\to \left\{R_U,\infty\right\}$. For asymptotically dS space, this is not allowed since $r$ is limited by the cosmic horizon radius $R_{U}=\sqrt{(D-1)(D-2)/2\Lambda_{D}}$. For asymptotically AdS space, $R\to \infty$ is physically viable. Generically, the bounds given by Eq. (\ref{ubounds}) may yield bounds on the average density of static, asymptotically dS and AdS universes by letting $R=R_{U}$ and $R\to \infty$, respectively.

First, let us consider the AdS case. Since $\Gamma$, given by Eq.~(\ref{Gam}), is proportional to $R^{3}/y$, dividing by $R^{2}$ gives an interesting constraint on the average density of a {\it static} universe, with $D>3$. For $R\to \infty$, the parameters in Eq. (\ref{uform}) become
\begin{eqnarray}
\frac{B}{R^{2}} &\to& 2(D-1)\left( \frac{\Lambda_{D}}{D-2}(1+w)-\frac{\Gamma}{R^{2}}\right), \nonumber \\
\frac{C}{B^{2}}&=& \frac{1}{4(D-1)^{2}}=\frac{1}{4A}, \nonumber
\end{eqnarray}
leading to the degeneracy of $u_{+}$ and $u_{-}$. This implies the {\it uniqueness} of the average density of the static asymptotically AdS universe, which is given by
\begin{eqnarray}
\bar{\rho}^{\rm AdS}&=&\frac{1}{\kappa c^{2}}\left( \frac{\Gamma (D-2)}{R^{2}}-\Lambda_{D}(1+w)\right),
\end{eqnarray}
where
\begin{eqnarray}
\frac{\Gamma}{R^{2}}\approx \frac{2\kappa e^{2}}{\lambda_{e}^{2D-3}}\sqrt{\frac{(D-1)(D-2)}{-\Lambda_{D}-\kappa c^{2} \bar{\rho}^{\rm AdS}}}.
\end{eqnarray}
This bound only exists, for $\Lambda_{D}<-\kappa c^{2}\bar{\rho}^{\rm AdS}, \ w\gtrsim -1$, \emph{if} the universe is charged.
For the uncharged case, $Q=0$, the maximum mass/radius power ratio of the entire universe ($R\to \infty$) gives the average density bound
\begin{eqnarray}
\bar{\rho}^{\rm AdS}&=& -\frac{\Lambda_{D}c^{2}}{8\pi G}\left( 1+w \right),
\end{eqnarray}
for the {\it static} AdS universe.

For asymptotically dS space, we can set $R = R_{U}=\sqrt{(D-1)(D-2)/2\Lambda_{D}}$ to obtain an approximate average density
\begin{eqnarray}
\bar{\rho}^{\rm dS}&=& \frac{\Lambda_{D}}{\kappa c^{2}}\left( \frac{\Gamma (D-2)}{\Lambda_{D}R^{2}}-(1+w)+\frac{2(D-2)}{(D-1)^{2}}\right). \nonumber \\
\end{eqnarray}
For a static uncharged dS universe, starting from Eq. (\ref{olims}) and setting $R=R_{U}$, we obtain
\begin{eqnarray}
\bar{\rho}_{\rm max,min}^{\rm dS}&=& \frac{\Lambda_{D}}{\kappa c^{2}}\Bigg( 1+w+\frac{2(D-2)}{(D-1)^{2}}\nonumber \\
&&\pm \frac{2(D-2)}{(D-1)^{2}}\sqrt{1+w\frac{(D-1)^{2}}{(D-2)^{2}}}~ \Bigg).
\end{eqnarray}
These limits exist only for $w>-(D-2)^{2}/(D-1)^{2}$. Even if, in reality, the universe is an expanding $(3+1)$-dimensional space-time with positive $\Lambda$, the static dS and AdS space still have essential interpretations from the viewpoint of holographic duality (c.f. discussion of the AdS case in Sec. \ref{Sec.4}).

\section{Total energy and gravitational stability of charged objects with
minimum mass/radius ratio in arbitrary dimensions}  \label{Sec.6}

In the present Section, we investigate the stability of charged
gravitating objects in arbitrary dimensions. As a first step in this study,
we derive an explicit expression for the total energy of a compact, charged,
general relativistic object, which includes the contribution from the
gravitational energy. For the sake of notational convenience, we use a system of units such
that $c=G=1$ and $\kappa =8\pi$ throughout the remainder of this Section.

A definition of gravitational field energy $E_G$, with interesting
properties, was proposed in \cite{LyKa85}, and further developed in \cite%
{Gr86,Gr92}. The derivation of $E_G$ proceeds as follows: Let us assume that $T_{\nu }^{\mu }$ is
the energy-momentum tensor of a stationary system with mass $M$, embedded within a space-time with a
time-like Killing vector $\xi^{\nu}$. Then the matter energy $E_{M}$ of the
system is defined as \cite{LyKa85,Gr86}
\begin{equation}
E_{M}=\int T_{\nu }^{\mu }\xi ^{\nu }\sqrt{-g}d\Sigma _{\mu }\text{,}
\end{equation}%
where $\Sigma $ is any space-like surface over which the energy is to be
evaluated. If $M$ is the total energy of the system, then the gravitational
energy of the system $E_{G}$ is defined as $E_{G}=M-E_{M}$ \cite{LyKa85}.

This definition of the gravitational energy can be reformulated in terms of
the theory of surface layers \cite{Gr86,Gr92} as follows: Let the surface $\Sigma $ be
a closed surface, which cuts the space-time in such a way that the exterior
space-time remains unchanged, while the interior space-time is flat.
The internal energies of the matter and of the gravitational field are then
replaced by the surface energy of $\Sigma $, so that $E_{G}=E_{\Sigma }-E_{T}
$, where $E_{G}$ is the energy of the gravitational field inside $\Sigma $, $%
E_{\Sigma }$ is the energy of $\Sigma $, and $E_{T}$ is the energy of the matter
inside $\Sigma $ \cite{Gr86}. The matter energy inside the surface is given
by $E_{T}=\int_{V}T_{\nu }^{\mu }\xi ^{\nu }u_{\mu }dV$, where $V$ is the
invariant volume inside $\Sigma $, and $u_{\mu }$ is the four-velocity field
of points that are fixed in $\Sigma $.

Next, we introduce the unit normal
vector field $\mathbf{n}$ of $\Sigma $. The exterior curvature tensor of $%
\Sigma $ is defined by $K_{ij}=\nabla _{j}n_{i}$, where we have introduced
the set of intrinsic coordinates $\left( x^{i},x^{j}\right) $ on $\Sigma $.
The surface energy tensor of $\Sigma $ (the Lanczos tensor) $S_{j}^{i}$ is
defined as $S_{j}^{i}=\left( 1/8\pi \right) \left( \left[ K_{j}^{i}\right]
-\delta _{j}^{i}\left[ K\right] \right) $, where $\left[ {}\right] $ denotes
the discontinuity at $\Sigma $, and $K=K_{i}^{i}$ \cite{Gr86}. The energy of
the cut is given by $E_{\Sigma }=\int_{\Sigma }S_{\nu }^{\mu }\xi ^{\nu
}u_{\mu }d\Sigma $, where $\Sigma $ is an invariant surface element. For a
vacuum solution of Einstein's field equations, $E_{\Sigma }$ gives the
gravitational field energy inside $\Sigma $. If there is a nonvanishing
energy-momentum density tensor inside $\Sigma $, then the gravitational
field energy is given by \cite{Gr86}
\begin{equation}
E_{G}=\int_{\Sigma }S_{\nu }^{\mu }\xi ^{\nu }u_{\mu }d\Sigma
-\int_{V}T_{\nu }^{\mu }\xi ^{\nu }u_{\mu }dV.
\end{equation}%
This definition is manifestly coordinate invariant.

In the case of spherical symmetry, and in arbitrary dimensions, $%
S_{0}^{0}=-\left( c^{4}/4\pi G\right) \sum_{i=1}^{D-2}\left[ K_{\theta
_{i}}^{\theta _{i}}\right] $, and thus the energy inside the surface $\Sigma $
is given by
\begin{eqnarray}
E_{G} &=&-(D-3)re^{\nu /2}\left[ e^{-\lambda /2}\right] =  \notag \\
&&-(D-3)re^{\nu /2}\left( 1-e^{-\lambda /2}\right) .
\end{eqnarray}
In the exterior of a higher-dimensional charged matter distribution, the vacuum
metric functions satisfy the condition $\nu +\lambda =0$, and the metric is
the generalized $D$-dimensional Reissner-Nordstrom-de Sitter metric, with
coefficients
\begin{eqnarray}
e^{\nu } &=&e^{-\lambda }=1-2u+\Delta -\frac{2 b}{D-1},  \notag \\
&&
\end{eqnarray}
where the dimensionless parameters $u$, $\Delta$ and $b$ are given by Eq.~(\ref{dlpar}).
Therefore, the gravitational energy inside a
surface of radius $R$, where $R$ is the radius of the charged object, is
given by
\begin{widetext}
\begin{eqnarray}
\left.E_{G}\right |_{r=R} &=&-(D-3)R\sqrt{1-\frac{2MF(D)}{R^{D-3}}+%
\frac{\kappa Q^{2}}{\left( D-2\right) \left( D-3\right) }\frac{1}{R^{2\left(
D-3\right) }}-\frac{2 \Lambda _{D}}{\left( D-1\right) \left( D-2\right)
}R^{2}} \nonumber\\
&&\left[ 1-\sqrt{1-\frac{2MF(D)}{R^{D-3}}+%
\frac{\kappa Q^{2}}{\left( D-2\right) \left( D-3\right) }\frac{1}{R^{2\left(
D-3\right) }}-\frac{2 \Lambda _{D}}{\left( D-1\right) \left( D-2\right)
}R^{2}}%
\right] ,
\end{eqnarray}
\end{widetext}
where $F(D)\equiv \kappa/[(D-2)\Omega_{D-2}]$. For an object with minimum mass/radius power ratio, and with small $\Delta, b$ and negligible $\Gamma$, the condition (\ref{mrmin}) is satisfied. Thus, eliminating the total mass using Eq. (\ref{mrmin}), the gravitational energy becomes
\begin{widetext}
\begin{eqnarray}
\left.E_{G}\right |_{r=R} &=&-(D-3)R\sqrt{1-
\frac{\kappa Q^{2}}{\left( D-2\right)^{2}}\frac{1}{R^{2\left(
D-3\right) }}+\frac{2 w \Lambda _{D}}{\left( D-2\right)^{2}
}R^{2}}\left[ 1-\sqrt{1-
\frac{\kappa Q^{2}}{\left( D-2\right)^{2}}\frac{1}{R^{2\left(
D-3\right) }}+\frac{2 w \Lambda _{D}}{\left( D-2\right)^{2}
}R^{2}} ~\right] .   \notag \\
\end{eqnarray}
\end{widetext}

For a stable configuration, the total gravitational energy
should have a minimum, defined by
\begin{equation}
\frac{\partial E_{G}}{\partial R}=0, \ \ \ \frac{\partial^2 E_{G}}{\partial R^2} > 0.
\end{equation}%
The resulting expression can be rearranged into a cubic equation in $\Lambda_D$, and thus solved analytically for $\Lambda_{D}$ in arbitrary dimensions.  However, the expression is long and complicated, so we will give only the approximate form, valid to leading order in $Q^{2}$ and $\Lambda_D$. Thus, we have
\begin{eqnarray}\label{stcon}
\frac{(2D-7)}{2}R^{2(3-D)}\kappa Q^{2}  
+3\Lambda_{D}wR^{2} \approx 0,
\end{eqnarray}
giving
\begin{eqnarray}\label{radius}
R & \approx & \left( \frac{2D-7}{6}\frac{\kappa Q^{2}}{-w\Lambda_{D}}\right)^{1/[2(D-2)]},
\end{eqnarray}
as the radius of a stable charged object, with minimum mass/radius power ratio, which also minimizes the gravitational energy of the configuration.

Following the method presented in~\cite{Boehmer:2007gq}, for compact objects in $(3+1)$ dimensions, we obtain the minimum mass of a charged object as a function of the $D-$dimensional cosmological constant $\Lambda_{D}$, and of its electric charge, in the form
\begin{eqnarray}  \label{massstab}
M&=&g(w,D)\frac{\Omega_{D-2}}{D-2}Q^{\frac{D-1}{D-2}}\left( \frac{-6w\Lambda_{D}}{(2D-7)\kappa} \right)^{\frac{D-3}{2(D-2)}}
\end{eqnarray}
where
\begin{equation}
g(w,D)\equiv \frac{2D-5}{2(D-3)}+\frac{2D-7}{6}\left( \frac{D-2}{w(D-1)}+1\right).
\end{equation}
Using Eqs.~(\ref{radius})-(\ref{massstab}), we can now eliminate the cosmological constant and obtain the ratio of the square of the charge of the object to its mass, as a function of radius, as
\begin{equation}\label{QM}
\frac{Q^2}{M}=\frac{D-2}{g(w,D)\Omega_{D-2}}R^{D-3}.
\end{equation}
For $D=4$, this gives
\begin{eqnarray} \label{QM}
\frac{Q^2}{M}=\frac{1}{4\pi}\left(\frac{18w}{1+15w}\right)R
= \left(\frac{1}{4\pi}\right)\frac{9}{7}R,~(w=-1).
\end{eqnarray}
From the expression for $R$ in Eq.~(\ref{radius}), we see that stability can be achieved only when $\Lambda_{D}>0$ for negative $w$. Furthermore, if we fix the mass $M$ the of object to the minimum allowed at a given radius $R$, and consider a change $\delta R$, then any change in the charge contribution $Q^{2}$ must be compensated by a corresponding change in the $\Lambda_{D}$ contribution, in such a way that as to keep the gravitational energy of the system constant.

In the case of the electron, for which $Q=e$ 
and $M=m_e$, Eq.~(\ref{QM}) automatically recovers the classical electron radius, $r_e=1.28 e^2/m_e$~(in CGS units). In classical, relativistic, but \emph{non-gravitational} physics, this is obtained from the requirement that the electrostatic energy $e^2/r_e$ equals the rest-mass energy $m_e$. In the present approach, this result is obtained by minimizing the {\it total gravitational energy} of a charged system with minimum mass/radius ratio.

\subsection{A complementary stability analysis} \label{Sec.6.1}

Consider a charged static object with the minimum mass/radius power ratio.  One way to obtain a stability condition for this object is by minimizing the quantity $M/R^{D-3}_{\rm min}$ with respect to the radius $R$.   If the object has the minimum mass/radius power ratio and we make a slight change in $R$, its mass has to change in such a way that the mass/radius power ratio remains constant.  By setting
\begin{equation}
\frac{\partial}{\partial R}\left( \frac{M}{R^{D-3}}\right)_{\rm min}=0,
\end{equation}
and using condition (\ref{mrmin}), we obtain the stability condition
\begin{equation}
\frac{\kappa Q^{2}}{R^{2(D-2)}} = -\frac{2\Lambda_{D}}{2D-5} \left( \frac{D-2}{D-1}+w \right),
\end{equation}
yielding the radius
\begin{equation}\label{rD}
R = \left[ \frac{\kappa Q^{2}(2D-5)}{-2\Lambda_{D}\left( \frac{D-2}{D-1}+w \right)} \right]^{\frac{1}{2(D-2)}}.
\end{equation}
Substituting back into Eq. (\ref{mrmin}) gives the mass at the minimum mass/radius power ratio configuration,
\begin{equation}\label{mD}
M = \frac{\Omega_{D-2}(2D-5)~Q^{2}}{2(D-3)}\left[ \frac{\kappa Q^{2}(2D-5)}{-2\Lambda_{D} \left( \frac{D-2}{D-1}+w \right)} \right]^{-\frac{D-3}{2(D-2)}}.
\end{equation}
The dark energy constant $\Lambda_{D}$ can be eliminated using Eqs. (\ref{rD})-(\ref{mD}) to give
\begin{equation}\label{QMD}
\frac{Q^2}{M}=\frac{2(D-3)}{(2D-5)\Omega_{D-2}}R^{D-3}.
\end{equation}
For $D=4$, this gives
\begin{equation}\label{QM4}
\frac{Q^2}{M}=\frac{1}{6\pi}R^{D-3}.
\end{equation}

Although the stability analysis based on the minimization of gravitational energy allows only positive $\Lambda_{D}$, for negative $w$, the stability analysis presented here allows both $\Lambda_{D} > 0$ and $\Lambda_{D} < 0$. However, for $\Lambda_{D} <0$, the equation of state parameter $w$ must satisfy the condition $w>-(D-1)/(D-2)$, which gives $w > -3/2$ for $D=4$.

\section{Quantum implications of a classical minimum mass for charged objects} \label{Sec.7}

In this Section, we investigate the quantum implications
of the existence of a classical minimum mass for
charged objects in $(3 + 1)$-dimensions. In Sec. \ref{Sec.7.1}, we
begin by reviewing a series of quantum gravity arguments
that give rise to `cubic' MLURs, in which the minimum positional uncertainty $(\Delta x)_{\rm min}$ is given by the cube root
of three phenomenologically significant length scales. In general, such relations may be derived by minimizing the total uncertainty,
due to both canonical quantum mechanical and gravitational effects, with respect to the mass $M$ of the system.
In Sec. \ref{Sec.7.2}, we combine the mass minimization condition
giving rise to cubic MLURs with phenomenological
results from canonical quantum mechanics, namely, the existence of a minimum Compton radius for any object (i.e. `particle') of mass $M$, and consider
charged objects subject to the bound (\ref{R_class}).
By combining all three
mass bounds, we obtain the condition for quantum gravitational stability of a charged particle. Applying this to the electron, we find that saturation of this condition
\emph{requires} the existence of a `new' fundamental length scale in nature, $R_*$, of order $R_W$. Furthermore, setting $R_* = R_W = \sqrt{3/\Lambda}$, we recover the expression
for $\Lambda$, written in terms of the fundamental constants $\left\{c,G,\hbar,e,m_e\right\}$, Eq. (\ref{Lambda}).

For later convenience, we now define the Planck length $R_P$,
mass $M_P$, and charge $q_P$, expressed in terms of the independent constants
$\left\{c,G,\hbar\right\}$, via
\begin{eqnarray}  \label{Planck-1}
R_P = \sqrt{\frac{\hbar G}{c^3}}, \ \ \ M_P = \sqrt{\frac{\hbar c}{G}},
\end{eqnarray}
and
\begin{eqnarray}  \label{q_P}
q_P = \sqrt{\hbar c}.
\end{eqnarray}
For the sake of clarity, \emph{all} fundamental constants are written explicitly
throughout the remainder of this Section.

Following \cite{Wesson:2003qn}, but adopting the notation and terminology
used in \cite{Burikham:2015nma}, we also define two mass scales, $M_W$ and $%
M_W^{\prime }$, associated with the cosmological constant $\Lambda$,
\begin{eqnarray}  \label{Wesson-M}
M_W = \frac{\hbar}{c}\sqrt{\frac{\Lambda}{3}}, \ \ \ M^{\prime }_W = \frac{%
c^2}{G}\sqrt{\frac{3}{\Lambda}}.
\end{eqnarray}
From here on, we refer to these as the first and second Wesson masses,
respectively. The associated lengths are
\begin{eqnarray}  \label{Wesson-M}
R_W = \sqrt{\frac{3}{\Lambda}}, \ \ \ R^{\prime }_W = \frac{\hbar G}{c^3}%
\sqrt{\frac{\Lambda}{3}}.
\end{eqnarray}
which we will refer to as the first and second Wesson length scales. Hence, $%
R_W$ is simply the Compton wavelength of a particle of mass $M_W$ and $%
R_W^{\prime }$ is the Compton wavelength of a particle of mass $M^{\prime }_W
$. We also note that $M^{\prime }_W = M_P^2/M_W$ and $R^{\prime }_W =
R_P^2/R_W$, so that $\left\{M_W,R_W\right\} \leftrightarrow \left\{M^{\prime
}_W,R^{\prime }_W\right\}$ under the $T$-duality transformations (see, for
example \cite{Alvarez:1994dn}), $M_W \leftrightarrow M_P^2/M_W$, $R_W
\leftrightarrow R_P^2/R_W$. Physically, $M^{\prime }_W \approx 1.347 \times
10^{56}$ g and $R_W \approx 1.0 \times 10^{28}$ cm are of the order of the
present day mass and radius of the Universe \cite{Burikham:2015nma}.

\subsection{Cubic MLURs in phenomenological quantum gravity} \label{Sec.7.1}

In addition to those proposed by Bronstein \cite{Bronstein}, at least three
sets of heuristic arguments based on quantum gravitational phenomenology
give rise to the cubic MLURs. The first is based on an extension of a gedanken
experiment first proposed by Salecker and Wigner \cite{Salecker:1957be}, which
proceeds as follows. Suppose we attempt to measure a length $d$ using a
special `clock', consisting of a mirror and a device that both emits and
detects photons. The photons are reflected by the mirror, placed at some
\emph{unknown} length $d$ from the device, which emits a photon and
reabsorbs it after a time $t = 2d/c$. Assuming that the recoil velocity of the device is well below the speed of light,
it may be modeled non-relativistically. Then, by the standard HUP,
the uncertainty in its velocity $\Delta v$, at the time of emission, is of
order
\begin{eqnarray}  \label{dv}
\Delta v = \frac{\hbar}{2M \Delta x},
\end{eqnarray}
where $M$ is its mass and $\Delta x$ is the \emph{initial} uncertainty in
its position.

We note that the `device' considered here may still be small enough to behave quantum mechanically.
For example, we may consider a two-state system involving a charged particle, embedded within a broader experimental set-up, that emits and re-absorbs photons.
In this case, $\Delta x$ and $\Delta p = M\Delta v$ refer to the positional and momentum uncertainty of the charged particle, which, together with the mirror that reflects the photons, measures (or `probes') the distance $d$.

During the time required for the photon to travel to the
mirror and back, the particle acquires an additional positional uncertainty $%
(\Delta x)^{\prime }= 2d\Delta v/c$ (i.e., in addition to the standard
positional uncertainty $\Delta x \gtrsim \hbar/(2 M\Delta v)$), so that the
\emph{total} positional uncertainty is given by
\begin{eqnarray}  \label{dx_tot-1}
\Delta x_{\rm tot} = \Delta x + \frac{\hbar d}{Mc \Delta x} = \frac{\hbar}{2 M
\Delta v} + \frac{2d\Delta v}{c}.
\end{eqnarray}
Minimizing this expression with respect to $\Delta x$, and using $(\Delta v)_{\rm max} \approx  \hbar/(M(\Delta x)_{\rm min})$, we have
\begin{eqnarray}  \label{Wigner-2}
(\Delta x)_{\rm min} &\approx & \sqrt{\frac{\hbar d}{Mc}} \approx \sqrt{\lambda_C d},
\notag \\
(\Delta v)_{\rm max} &\approx & \sqrt{\frac{\hbar c}{Md}} \approx c\sqrt{\frac{%
\lambda_C}{d}},
\end{eqnarray}
(neglecting numerical factors of order unity), where $\lambda_C = \hbar/(Mc)$ denotes the Compton wavelength of the particle, so that
\begin{eqnarray}  \label{dx_tot_min}
(\Delta x_{\rm tot})_{\rm min} \approx \sqrt{\lambda_C d}.
\end{eqnarray}
If we then require $d > R_S = 2GM/c^2$ (i.e. that our measuring device is
not inside a black hole), we obtain $(\Delta x_{\rm tot})_{\rm min} = R_P$. However, more realistically,
we may require $d > \lambda_C$, so that the measurement process devised by Salecker and Wigner gives rise to a MLUR
which is consistent with the standard Compton bound.

The original argument presented in \cite{Salecker:1957be} may also be
modified to explicitly include the \emph{classical} `uncertainty' in the position of the
measuring device due to gravitational effects. Assuming that this is
proportional to the Schwarzschild radius of the device $R_S$, the total
uncertainty due to canonical quantum mechanical effects, plus gravity, is
\begin{eqnarray}  \label{dx_tot-2}
\Delta x = \sqrt{\frac{\hbar d}{Mc}} + \beta\frac{GM}{c^2},
\end{eqnarray}
where $\beta >0$ \cite{Ng:1994zk}. Minimizing this with respect to $M$
yields
\begin{eqnarray}  \label{M_min-1}
M \simeq M_P\left(\frac{d}{\beta^2R_P}\right)^{1/3}
\end{eqnarray}
and, substituting this back into Eq. (\ref{dx_tot-2}), we obtain
\begin{eqnarray}  \label{MLUR-1}
\Delta x \geq (\Delta x)_{\rm min} \simeq (\beta R_P^2d)^{1/3},
\end{eqnarray}
where (again) we have neglected numerical factors of order unity in the preceding expressions.

One disadvantage of the approach described above is that it appears to
apply only to the specific measurement process envisaged in \cite%
{Salecker:1957be}. However, in \cite{Calmet:2004mp,Calmet:2005mh}, it was
shown that the expression for $(\Delta x)_{\rm min}$ given in Eq. (\ref{Wigner-2}%
) may be obtained from general principles in canonical quantum mechanics. For
$V=0$, the time evolution of the position operator $\hat{x}(t)$ given by the
Schr{\" o}dinger equation (in the Heisenberg picture) is
\begin{eqnarray}  \label{Heis-1}
\frac{d\hat{x}(t)}{dt} = \frac{i}{\hbar}[\hat{H},\hat{x}(t)] = \frac{\hat{p}%
}{M}.
\end{eqnarray}
This may be solved directly to give
\begin{eqnarray}  \label{Heis-2}
\hat{x}(t) = \hat{x}(0) + \hat{p}(0)\frac{t}{M}.
\end{eqnarray}
The spectra of any two Hermitian operators, $\hat{A}$ and $\hat{B}$, must
obey the general uncertainty relation \cite{Isham}
\begin{eqnarray}  \label{AB}
\Delta A \Delta B \geq \frac{1}{2}|\langle[\hat{A},\hat{B}]\rangle|,
\end{eqnarray}
so that setting $\hat{A} = \hat{x}(0)$, $\hat{B} = \hat{x}(t)$ gives
\begin{eqnarray}
[\hat{x}(0),\hat{x}(t)] = i\hbar \frac{t}{M},
\end{eqnarray}
and
\begin{eqnarray}
\Delta x(0)\Delta x(t) = \frac{\hbar t}{2M}.
\end{eqnarray}
In the Heisenberg picture, we have $(\Delta x)^2 = \Delta x(0)\Delta x(t)$,
so that, again setting $t = d/c$, $(\Delta x)_{\rm min}$ is given by Eq. (\ref%
{Wigner-2}).

As with Salecker and Wigner's gedanken experiment, we have again considered
performing two measurements of the position of an object, one at $t=0$ and
the other at some time $t > 0$, and can relate this to the uncertainty
inherent in the measurement of a length scale $d = ct$. However, in this
case, no assumptions have been made about the details of the measurement
procedure, so that Eq. (\ref{Wigner-2}) may be considered a general result
in canonical quantum mechanics (i.e. not accounting for the effects of gravity).
As such, the arguments presented in \cite{Ng:1994zk}, and hence the cubic
MLUR (\ref{MLUR-1}), may be considered to have general validity for
gravitating quantum mechanical systems.

Cubic MLURs of the form (\ref{MLUR-1}) (with $\beta = 1$) were also obtained in \cite{Karolyhazy:1966zz,KFL}
by considering a gedanken experiment to measure the lengths of geodesics with minimum quantum uncertainty.
This derivation relies on the fact that the mass of the measuring device $M$
distorts the background space-time. Equating the uncertainty in momentum of
the device with the uncertainty in its mass then implies an irremovable
uncertainty or `fuzziness' in the space-time in the vicinity of the device
itself. This results in an absolute minimum uncertainty in the precision
with which a \emph{gravitating} measuring device can measure the length of
any given world-line, $d$.

As with the results proposed in \cite{Ng:1994zk}, in this scenario the value $\beta \sim \mathcal{O}(1)$
arises as a direct result of the assumption that the Schwarzschild radius of a body of mass $M$, $R_S = 2GM/c^2$, represents the
minimum \emph{classical} `gravitational uncertainty' in its position. In fact, for cubic MLURs of the form (\ref{MLUR-1}), it is usually assumed that
$\beta \sim \mathcal{O}(1)$ in most of the existing quantum gravity literature \cite{Hossenfelder:2012jw}. For all of the scenarios
leading to Eq. (\ref{MLUR-1}) considered above, this is directly equivalent to assuming a minimum \emph{classical} gravitational uncertainty, given
by $R_S$.

However, since Eq. (\ref{MLUR-1}) holds if and only if Eq. (\ref{M_min-1}) also holds, it is
straightforward to check that setting $\beta = 1$ is  inconsistent with the
requirement that quantum gravity effects, stemming from MLURs of the form (\ref{MLUR-1}), be subdominant to `standard'
quantum effects. Since quantum gravity has not been observed in the lab,  we
require $(\Delta x)_{\rm min} \simeq (\beta R_P^2d)^{1/3} \leq \hbar/(Mc)$, or,
equivalently
\begin{eqnarray}  \label{KFL-3}
M \leq M_P\left(\frac{R_P}{\beta d}\right)^{1/3}.
\end{eqnarray}
Substituting the minimization condition for $\Delta x$, Eq. (\ref{M_min-1}),
into this inequality then gives
\begin{eqnarray}  \label{KFL-4}
d \leq \sqrt{\beta}R_P.
\end{eqnarray}
Clearly, for $\beta \sim \mathcal{O}(1)$, this contradicts the weak
gravitational limit of the theory, represented by Eq. (\ref{Wigner-2}), and
which yields $d \geq (\Delta x)_{\rm min} = R_P$. This implies that the
arguments of K{\' a}rolyh{\' a}zy \emph{et al} \cite{Karolyhazy:1966zz,KFL},
which automatically assume $\beta = 1$, are also inconsistent with the weak field limit
and the condition that quantum gravity effects from MLURs become subdominant to
canonical quantum uncertainty in this regime.

In fact, subsequent work has claimed that K{\' a}rolyh{\' a}zy's quantum
space-time MLUR is incompatible with observations in yet another sense, in
that it implies a vacuum energy density of the order the neutron star
density \cite{Diosi:1993vy}. While it would be interesting to repeat the
arguments presented in \cite{Diosi:1993vy} using the more general relation,
Eq. (\ref{MLUR-1}), and to consider the value of
$\beta$ required to reduce the neutron star density to the observed vacuum
density, this task lies outside the scope of the present paper and is left
to future work. However, for now, we note that only the more general
relation (\ref{MLUR-1}), with $\beta \gg 1$, is compatible with current
observations.

Although the arguments presented in this subsection do not allow us to fix
the value of $\beta$, or even the minimum value of $\beta$ required for
consistency with the weak field limit, we note that they yield \emph{two} conditions on the mass of
the measuring device $M$ in relation to the distance to be measured $d$,
Eqs. (\ref{M_min-1}) and (\ref{KFL-3}),
and that these relations involve only a single free constant. In the next
subsection, we combine these with the condition relating the mass $M$ and
radius $R$ of a gravitationally stable charged body, and explicitly consider an object of charge $e$
and mass $m_{e}$ (i.e. an electron). In so doing, we see that the consistency of all three
relations implies the identification of fundamental constants given in Eq. (\ref{Lambda}).

\subsection{Quantum gravitational bounds for stable charged objects}\label{Sec.7.2}

One possible definition of the quantum gravity regime is the requirement that
the positional uncertainty of an object, due to combined canonical and quantum gravitational effects,
be greater than or equal to its classical radius, $(\Delta x)_{\rm min} \geq R$.
(This is essentially the inverse of the requirement for classicality, that
the macroscopic radius of an object be larger than its quantum uncertainty.)
Thus, the conditions
\begin{eqnarray}  \label{5-1}
\lambda_C \geq (\Delta x)_{\rm min} \geq R
\end{eqnarray}
correspond to a regime in which the `particle' behaves quantum mechanically
and gravitationally, but in which specific quantum gravitational effects are
subdominant the standard Compton uncertainty. In this regime, we may
therefore assume that
\begin{eqnarray}  \label{5-2}
(\Delta x)_{\rm min} = (\beta R_P^2d)^{1/3} = \gamma \lambda_C,
\end{eqnarray}
where $\gamma \leq 1$. Likewise, we may set
\begin{eqnarray}  \label{5-3}
\xi R = (\Delta x)_{\rm min} = (\beta R_P^2d)^{1/3},
\end{eqnarray}
where $\xi \geq 1$, if we expect the object to display no classical
behavior. Clearly,
\begin{eqnarray}  \label{5-3}
\gamma \leq \xi.
\end{eqnarray}
with equality holding if and only if $\gamma = \xi = 1$.

For convenience, we now rewrite the three independent expressions we have
obtained for $M$ throughout the preceding sections of this work, namely
\begin{subequations}
\begin{align}
M &= M_P\left(\frac{d}{\beta^2R_P}\right)^{1/3},  \label{M-1A} \\
M &= \frac{\gamma R_PM_P}{(\beta R_P^2d)^{1/3}},   \label{M-1B} \\
M &= \frac{Q^2}{q_P^2}\frac{\xi M_PR_P}{(\beta R_P^2d)^{1/3}}.  \label{M-1C}
\end{align}
Equations (\ref{M-1A}) and (\ref{M-1B}) are simply Eqs. (\ref{M_min-1}) and (%
\ref{5-2}) restated. Equation (\ref{M-1C}) corresponds to saturating the
bound in Eqs. (\ref{eqn:c})-(\ref{R_class}) by assuming that $R = (\Delta x)_{\rm min}/\xi$ represents
the value of the classical radius that minimizes the ratio $M/R$, for a sphere of mass $M$
and charge $Q$. (For simplicity, we have neglected numerical factors of
order unity in Eq.~(\ref{R_class})).

Thus, the quantity $M$ in equations (\ref{M-1A})-(\ref{M-1B}) denotes the value of the mass for which the quantum
uncertainty of the object, including gravitational effects, is minimized,
whereas the $M$ in Eq. (\ref{M-1C}) is the mass of a body for which the classical bound (\ref{R_class}) is saturated.
By Eq. (\ref{QMD}), this is also the radius at which both the classical mass/radius ratio and classical gravitational energy are both minimized.

We now investigate the properties of a charged particle for which the combined uncertainty (due to both canonical quantum and gravitational effects)
{\it and} the classical mass/radius ratio and gravitational energy minimised. Thus, we proceed by equating the three expressions for $M$ in Eqs. (\ref{M-1A})-(\ref{M-1C}).
The physical picture is that we use a `particle' of mass $M$ and classical radius $R$ as a probe to measure a distance $d$:
{\it the minimum uncertainty in the position of the particle is also the minimum
uncertainty in the measurement of $d$.}

Equations (\ref{M-1B}) and (\ref{M-1C}) immediately imply
\end{subequations}
\begin{eqnarray}  \label{gamma}
\frac{Q^2}{q_P^2} = \frac{\gamma}{\xi} \leq 1,
\end{eqnarray}
or, equivalently,
\begin{eqnarray}  \label{Q<q_P}
Q \leq q_P.
\end{eqnarray}
This gives a nice (and self-consistent) interpretation of the Planck charge $%
q_P$ as the leading order contribution to a sum of terms that determine the
maximum possible charge of a stable, gravitating, quantum mechanical object.
The bound (\ref{Q<q_P}) may also be obtained in a more direct way by
combining a general relativistic result with canonical quantum theory:
rewriting Eq. (\ref{M_min-1}) as $Q^2 \lesssim q_P^2 RM/(M_PR_P)$ and
invoking a Compton type relation between $R$ and $M$ (i.e. taking $R$ to be
the Compton radius of the sphere), yields exactly the same result.

For convenience, we now rewrite
\begin{eqnarray}  \label{R_*}
R_{*} = \beta d,
\end{eqnarray}
where $R_{*}$ is an arbitrary length scale. We note that, if $\beta$ is
independent of $d$, $R_{*}$ is simply proportional to $d$, but, if $\beta
\propto d^{-1}$, $R_{*}$ is a constant. (Also note that setting $\beta
\propto d^{-1}$ would in no way alter the argument for the minimization of $%
\Delta x$ with respect to $M$, proposed in Sec. \ref{Sec.7.1}.) Equations (%
\ref{M-1A}) and (\ref{M-1B})-(\ref{M-1C}) then become
\begin{eqnarray}
M = \frac{M_Pd}{(R_PR_{*}^2)^{1/3}},
\end{eqnarray}
and
\begin{eqnarray}  \label{M}
M = \gamma(M_P^2M_{*})^{1/3},
\end{eqnarray}
respectively, where $M_{*} = M_PR_P/R_{*}$. Equating the two then yields
\begin{eqnarray}  \label{d}
d = \gamma(R_P^2R_{*})^{1/3}.
\end{eqnarray}
The expression for $M$, Eq. (\ref{M}), also implies that the Compton
wavelength of the particle is given by
\begin{eqnarray}  \label{lambda_C}
\lambda_C = \frac{(R_P^2R_{*})^{1/3}}{\gamma} = \frac{\xi R}{\gamma} = \frac{%
q_P^2 R}{Q^2 \xi}.
\end{eqnarray}
The self-consistent solution to Eqs. (\ref{M-1A})-(\ref{M-1C}), written in
terms of $R_{*} = \beta d$ (\ref{R_*}), is therefore
\begin{eqnarray}  \label{gamma_xi}
\gamma = \frac{Q^2}{q_P^2} \leq 1, \ \ \ \xi = 1,
\end{eqnarray}
yielding
\begin{eqnarray}  \label{consistent-1}
R = \frac{Q^2}{q_P^2}\lambda_C = (R_P^2R_{*})^{1/3},
\end{eqnarray}
where $\lambda_C = \hbar/(Mc)$ and
\begin{eqnarray}  \label{consistent-2}
M = \frac{Q^2}{q_P^2}(M_P^2M_{*})^{1/3},
\end{eqnarray}
together with
\begin{eqnarray}  \label{consistent-3}
d = \frac{Q^2}{q_P^2}(R_P^2R_{*})^{1/3}.
\end{eqnarray}

To summarize our results, we have shown that, to probe a distance $d$, given
in terms of some length scale $R_{*}$ by Eq. (\ref{consistent-3}), we can
minimize the combined quantum mechanical and gravitational uncertainty inherent in the measurement
of $d$ by choosing an appropriate probe `particle', with mass $M$, given by Eq. (\ref%
{consistent-2}). This mass corresponds to a classical radius $R$ given by
Eq. (\ref{consistent-1}). In addition, we found that the minimum value of the combined
uncertainty, incorporating gravitational effects, is equal to the classical radius, $(\Delta
x)_{\rm min} = R$. (Alternatively, if we take a particle of mass $M$, given in terms of $Q$ and $M_* = M_PR_P/R_*$ by Eq. (\ref{consistent-2}),
the value of $d$ obtained in Eq. (\ref{consistent-3}) represents the length scale which is naturally `probed' by such a particle, with minimum uncertainty $(\Delta x)_{\rm min} = R$.)

Mathematically, the three length scales $\lambda_C$, $(\Delta x)_{\rm min} = R$, and $d$, are related via
\begin{eqnarray}  \label{consistent-4}
d = \frac{Q^2}{q_P^2}(\Delta x)_{\rm min} = \frac{Q^2}{q_P^2}R = \frac{Q^4}{q_P^4}\lambda_C.
\end{eqnarray}
We also showed that the Planck charge acts as a maximum possible charge for
any stable, gravitating, quantum mechanical object, regardless of its mass $M
$ and associated Compton radius $\lambda_C$. Therefore, for $Q < q_P$, the
positional uncertainty induced by quantum gravitational effects is strictly
less than the Compton scale, for a stable body of any mass $M$.

Let us now reverse this argument by asking the following question: If we
suppose that $M$ represents the mass, not of a composite body, but of a \emph{true}
fundamental particle, what is the inherent length scale that such a particle
can probe, with minimal uncertainty? To answer this question, we first note that, in order for $M$ and
$Q$ in Eq. (\ref{consistent-2}) to be constants of nature, $R_{*}$ must also
be a constant of nature.

As a test case, let us now consider the electron by setting $M = m_e$ and $Q =
e$, for which Eq. (\ref{consistent-4}) recovers the well known relation $r_e = \alpha \lambda_e$,
were $\alpha \approx 1/137$ is the fine structure constant.
For convenience, let us also associate $R_*$ with another universal
constant, which we denote $\Lambda_*$, via the relation
\begin{eqnarray}  \label{Lamba_*_defn}
\Lambda_{*} := \frac{3}{R_{*}^2}.
\end{eqnarray}
From Eq. (\ref{consistent-2}), we then have
\begin{eqnarray}  \label{Lambda_*}
\Lambda_{*} = 3\frac{m_{e}^6G^2}{\alpha^6\hbar^4}.
\end{eqnarray}
Evaluating this numerically gives
\begin{eqnarray}  \label{Lambda_*=Lambda}
\Lambda_{*} \approx 1.4\times 10^{-56} \ {\rm cm}^{-2},
\end{eqnarray}
and we recall that
\be
\Lambda \approx 3.0
\times 10^{-56} \ {\rm cm}^{-2},
\ee
is the value of the cosmological constant implied by various observations \cite%
{Ostriker:1995rn,Tegmark:2003ud,Tegmark:2000qy,Hazra:2014hma,Zunckel:2008ti}.
This strongly suggests the identification $\Lambda_{*} = \Lambda$, $R_{*} =
R_W$.

We stress, however, that this identification is \emph{not} based
purely, or even primarily, on a numerical coincidence. Rather, our
requirement that the total uncertainty $\Delta x$, incorporating canonical quantum and gravitational effects, be
minimized for all stable bodies, including fundamental particles, \emph{%
requires} the existence of a fundamental length scale in nature which is many orders
of magnitude larger than the Planck length. Specifically, the minimization
of the combined canonical quantum and gravitational uncertainty of the electron requires the
existence of a fundamental constant, with dimensions $L^{-2}$, of the form (%
\ref{Lambda_*}). Formally identifying $\Lambda_{*} = \Lambda$ and
substituting this back into Eq. (\ref{eqn:c}), we obtain the bound
\begin{eqnarray}  \label{BOUND}
\frac{Q^2}{M} \lesssim \left(\frac{3\hbar ^2G^2c^6}{\Lambda}\right)^{1/6}  \approx \frac{e^2}{m_e} = 2.52 \times 10^{8} \;{\rm Fr/g},
\end{eqnarray}
to leading order. The fulfillment of this condition therefore
indicates the stability of a general, charged, gravitating, quantum
mechanical object, as claimed.

Finally, let us now consider the role of the length scale $d$, given by Eq. (%
\ref{consistent-3}), when $R_{*}$ is a universal constant and so is $Q$.
Combining this with the requirement that $R_* = \beta d$,
Eq. (\ref{R_*}), we obtain the following expression for $\beta$:
\begin{eqnarray}  \label{beta}
\beta = \frac{q_P^2}{Q^2}\left(\frac{R_*}{R_P}\right)^{2/3}=\frac{\lambda_{C}R}{R_{P}^{2}}.
\end{eqnarray}
The preceding arguments, given in Sec. \ref{Sec.7.1}, then imply that the
gravitational uncertainty of the particle is given by
\begin{eqnarray}  \label{grav_uncert}
(\Delta x)_{\rm grav} \approx \frac{\lambda_{C}R}{R_{P}^{2}} \times R_S = R = (\Delta x)_{\rm min},
\end{eqnarray}
rather than $(\Delta x)_{\rm grav} \approx R_S = 2GM/c^2$, as assumed in
\cite{Ng:1994zk,KFL,Diosi:1993vy}. This implies that an additional, self-consistent interpretation of
the classical radius $R$ of a charged object, is that it represents the minimum value of
the \emph{classical} gravitational `disturbance' induced by the objects mass $M$.

As an additional check on the consistency of this result, we note that
imposing the general condition $\beta = R_{*}/d$ on Eq. (\ref{dx_tot-2})
yields
\begin{eqnarray}  \label{X-1}
\Delta x = \sqrt{\frac{M_PR_P d}{M}} + \frac{R_{*}R_PM}{M_P d}.
\end{eqnarray}
Clearly, minimizing this expression with respect to $M$ (i.e. treating $M$
and $d$ as independent variables), yields
\begin{eqnarray}  \label{X-2}
(\Delta x)_{\rm min} = (R_P^2R_{*})^{1/3}.
\end{eqnarray}
However, since the length scale that may be probed with minimum total uncertainty using a
particle of canonical quantum width $\lambda_C$ is $d = (Q^4/q_P^4)\lambda_C$ (Eq. \ref{consistent-4}), where $Q^2 \leq q_P^2$, it is reasonable to ask,
what happens in the `canonical quantum limit', where $d \rightarrow \lambda_C$, so that $d$ and $M$ can no longer be
considered independent variables? In this case, for charged particle, Eq. (\ref{consistent-4}) requires $Q^2 = q_P^2$, and Eq. (\ref{X-1}) becomes
\begin{eqnarray}  \label{X-3}
\Delta x = d + \frac{R_{*}R_P^2}{d^2} = \frac{R_PM_P}{M} + \frac{R_{*}M^2}{%
M_P^2}.
\end{eqnarray}
Minimizing this expression with respect to either $d$ \emph{or} $M$ yields
Eq. (\ref{X-2}). Hence, Eq. (\ref{X-1}) is the \emph{only} form of Eq. (\ref%
{dx_tot-2}) which is valid, for charged particles, in the canonical quantum regime. However, we may conjecture that Eq. (\ref{X-3}) is valid in general, even for uncharged particles,
since the expression (\ref{dx_tot-2}) breaks down, for $\beta \sim \mathcal{O}(1)$, when particles are used to probe distances comparable to their canonical quantum radius.

We also note that, in the limit $d \rightarrow \lambda_C$, Eq. (\ref{beta}) implies $R_* \rightarrow \lambda_C^3/R_P^2$.
Substituting these values into Eq. (\ref{X-1}) yields
\begin{eqnarray}  \label{X-4}
\Delta x \approx (\Delta x)_{\rm min} = R = d = \lambda_C,
\end{eqnarray}
and
\begin{eqnarray}  \label{X-5}
R_* = \frac{\lambda_C^3}{R_P^{2}}.
\end{eqnarray}
Since we require $\lambda_C \gtrsim R_P$, in order to avoid black hole formation, all quantities $\left\{\Delta x,R ,d,R_*,\lambda_C\right\}$ remain above the Planck scale in this limit.

Finally, before concluding this Section, we note that Bronstein's bound (\ref{Bronstein-1}) is also a form of cubic MLUR, which may be rewritten as
\begin{eqnarray}  \label{X-5}
(\Delta x)_{\rm min} \approx \left(\frac{M_P^2}{M^2}\right)^{1/3}R,
\end{eqnarray}
after identifying $\Delta x \approx c\Delta t$, where we have used the fact that $\rho = MR^{-3}$ denotes the \emph{classical} density.
It is clear that this is compatible with our result, Eq. (\ref{consistent-4}), only when $M = M_P$. Thus, in general, our results are \emph{incompatible} with those presented in \cite{Bronstein}.

Although Bronstein did not explicitly consider charged particles, so that our results are not \emph{directly}
comparable to his, the origin of the difference appears to lie the fact that his results imply the gravitational field of an object gives rise to an additional uncertainty in its momentum, of order
$(\Delta p)_{\rm grav} \approx G\rho^2V\Delta x\Delta t$. Though it is beyond the scope of the present work to investigate this discrepancy further, it would be interesting to consider extending Bronstein's
original arguments to the case of charged bodies, to see whether they are compatible with those presented here.

\subsection{MLUR and Holography in arbitrary dimensions} \label{Sec.8}

In this Section, we will demonstrate that the MLUR which represents the minimum possible uncertainty due to combined canonical quantum and gravitational effects, implies holography involving quantum gravity `bits', in space-times with an arbitrary number of non-compact dimensions.

We have seen, in Sec.~\ref{Sec.7.1}, that the minimum canonical quantum mechanical uncertainty is proportional to $M^{-1/2}$, where $M$ is the mass of the object (c.f. Eq. (\ref{X-1})). It was also shown in Sec.~\ref{Sec.4} that, for a given mass of a static object, its classical radius has a minimal possible value before the object collapses to form a black hole. (This is because the ratio $M/R^{D-3}$ has an upper bound, see also \cite{Burikham:2015nma}). Thus, the minimum classical gravitational uncertainty is given by the minimum radius of the object, which is roughly the same as the horizon radius of the corresponding black hole.
This is proportional to $M^{1/(D-3)}$ in $D$ dimensions.

Therefore, the MLUR of the $D$-dimensional space-time can be expressed in the form
\begin{eqnarray}
\Delta x &\geq& \frac{\zeta}{\sqrt{M}}+\chi M^{1/(D-3)},
\end{eqnarray}
where $\zeta$, $\chi$ are positive constants. This expression contains the canonical quantum mechanical term and the classical gravitational term. By minimizing $\Delta x$ with respect to $M$, we obtain the minimum length
\begin{eqnarray}
(\Delta x)_{\rm min}&=&\frac{D-1}{2}\left( \frac{2}{D-3}\right)^{\frac{D-3}{D-1}}(\zeta^{2}\chi^{D-3})^{1/(D-1)},\nonumber \\
\end{eqnarray}
which corresponds to the mass
\begin{eqnarray}
M&=&\left( \frac{\zeta(D-3)}{2\chi} \right)^{2(D-3)/(D-1)}.
\end{eqnarray}
For a measuring apparatus with size $\ell$, the quantum mechanical uncertainty term and the classical gravitational term have parameters
\begin{eqnarray}
\zeta & \sim & \sqrt{\frac{\ell \hbar}{c}}, \nonumber \\
\chi & \sim & (\kappa c^{2})^{1/(D-3)}.
\end{eqnarray}
By using the $D$-dimensional Planck length
\begin{eqnarray}
R_{P(D)}&\sim& \left( \frac{\hbar\kappa c^{2}}{c}\right)^{1/(D-2)},
\end{eqnarray}
we can express the maximum number of degrees of freedom in an $\ell^{D-1}$ volume as
\begin{eqnarray}
N = \left( \frac{\ell}{(\Delta x)_{\rm min}}\right)^{D-1}\sim 
\left( \frac{\ell}{R_{P(D)}}\right)^{D-2}.
\end{eqnarray}
Remarkably, the result satisfies a holographic relation.

Thus, the maximum number of degrees of freedom in a $(D-1)$-dimensional volume is proportional to the $(D-2)$-dimensional `area' of the boundary in which the volume is enclosed. Specifically, it is equal to the number of quantum gravity bits, $(\ell/R_{P(D)})^{D-2}$, on the $(D-2)$-dimensional surface. Hence, we prove that cubic MLURs, combining the minimum possible uncertainties arising from both canonical quantum and gravitational effects, {\it inevitably} lead to holography in arbitrary, non-compact,
$D-$dimensional space-times.

\section{Discussions and final remarks} \label{Sec.9}

In the present work, we have investigated the possibility of the existence of a minimum mass/radius ratio for
charged, stable, compact general relativistic objects in arbitrary dimensions, in the presence of dark energy in the form of a cosmological constant. We have shown that for a static, spherically symmetric mass distribution, such a minimum ratio does indeed exist, and that it
arises as a direct consequence of the $D$-dimensional Buchdahl inequality, which also gives rise to an upper bound for the mass/radius ratio.

In the case of the minimum mass/radius ratio, we obtained an explicit inequality giving the lower bound on $M/R$ in arbitrary dimensions, as an explicit function of the charge $Q$ and the $D-$dimensional cosmological constant $\Lambda_D$. In order to obtain both the upper and lower bounds, we generalized the approach introduced in \cite{Burikham:2015nma} for uncharged objects to include nonzero charge, $Q \neq 0$. For $Q=0$, all our results reduce properly to the bounds obtained in \cite{Burikham:2015nma}.

In addition, we have investigated the condition of the thermodynamic stability for objects with minimum mass-radius ratio, which requires that they are in the minimum energy state. To estimate the total energy of these objects, we have use the definition of gravitational energy introduced in \cite{LyKa85}. In $D=4$ dimensions, imposing the condition of minimum stability, for charged objects with minimum mass/radius ratio, leads to an explicit expression, Eq.~(\ref{QM}), in which the ratio of the square of the charge of the object to its mass is proportional to the radius, $Q^2/M \propto R$. The same bound was also obtained as a stability condition for charged bodies in \cite{Boehmer:2007gq}.

We have also investigated the quantum implications of the existence of
a classical minimum mass for charged objects in four space-time dimensions, by starting from
 a series of quantum gravity arguments that give rise to cubic MLURs of the form
$\Delta x \geq (\Delta x)_{\rm min} = (\beta R_P^2d)^{1/3}$, Eq. (\ref{MLUR-1}), 
where $\beta$ is a positive numerical constant which is related to the positional uncertainty of the object induced by its gravitational field.
In these approaches, $\Delta x$
represents not only the uncertainty in the position of the object, but
also the irremovable quantum uncertainty inherent in any measurement of the
physical length $d$.

We have combined the mass minimization condition, giving rise to the cubic
MLURs, with phenomenological results from canonical quantum mechanics, namely, the existence of a minimum (canonical) quantum radius (the Compton radius), and have
considered objects subject to the minimum mass/radius bound for charged bodies (\ref{R_class}). By combining
all three mass bounds, we have obtained the condition for quantum gravitational
stability of a charged particle, $Q^2/M \lesssim \left(3\hbar^2G^2c^6/\Lambda\right)^{1/6}  \approx e^2/m_e = 2.52 \times 10^{8} \;{\rm Fr/g}$, Eq. (\ref{BOUND}).

Physically, we may interpret this as meaning that, if the electron were any less massive (for fixed charge $e$), or more highly charged (for fixed mass $m_e$), a combination of electrostatic and dark energy repulsion would lead to instability. In other words, the electron would blow itself apart, as claimed in the Introduction. In addition, saturation of this condition yields an expression for $\Lambda$ in terms
of the constants $\left\{c,\hbar,G,e,m_e\right\}$, given by Eq. (\ref{Lambda}).

Specifically, by combining the mass bound obtained from purely classical considerations with the cubic MLURs, motivated by quantum gravity, and applying the result to body of charge $e$ and mass $m_e$ (i.e. an electron), we obtained a prediction of a `new' constant of nature, $\Lambda_{*}$, which may be expressed in terms of other fundamental constants. Physically, the existence of this constant is \emph{required} in order to ensure the consistency of MLURs with the weak field limit of canonical quantum theory and with the classical stability bounds for charged, gravitating `particles'. Evaluating $\Lambda_{*}$ numerically, we have shown that it has the same order of magnitude value as the observed cosmological constant, which motivates the identification $\Lambda_{*} \equiv \Lambda \approx 10^{56}$ cm$^{-2}$. Crucially, this implies that, if the cosmological constant can be expressed as a function of the set of the `standard' constants $\left\{c,\hbar,G,e,m_e\right\}$, it cannot be interpreted as a {\it fundamental} constant of nature.

It is also interesting to note that the cubic MLURs used in this work, which, together with the classical stability bounds obtained for charged objects, imply a fundamental relationship between the existence of the cosmological constant and the stability of fundamental particles, \emph{also} imply the existence of a holographic relationship between the maximum number of degrees of freedom in a bulk space-time and the number of quantum gravity `bits' on the boundary. This is proved explicitly, for arbitrary $D-$dimensional space-times (with non-compact dimensions), in Sec. \ref{Sec.8}.

Finally, we note that the formalism developed in this paper can be easily extended to the case of non-electromagnetic interactions. For example, by interpreting the charge $Q$ as a generalised charge, corresponding to a Yang-Mills field, we can apply our results even to the case of strongly interacting particles, based on the fundamental QCD Lagrangian \cite{Wein}
\bea\label{eqYM}
L_{QCD}&=&\frac{1}{4}\sum_{a}F_{\mu \nu }^{a}F^{a\mu \nu }
\nonumber\\
&+&\sum_{f=1}^{N_{f}}%
\bar{\psi}\Big( i\gamma ^{\mu }\partial _{\mu }-\alpha _s\gamma ^{\mu }A_{\mu }^{a}%
\frac{\lambda ^{a}}{2}
-m_{f}\Big) \psi ,
\eea
where the subscript $f$ denotes the various quark flavors $u,d,s$ etc., and the corresponding quark masses $m_f$.  The
nonlinear gluon field strength is defined as
\begin{equation}\label{eqYM1}
F_{\mu \nu }^{a}=\partial _{\mu }A_{\nu }^{a}-\partial _{\nu }A_{\mu
}^{a}+\alpha _s f_{abc}A_{\mu }^{b}A_{\nu }^{c}.
\end{equation}
In Eqs.~(\ref{eqYM}) and (\ref{eqYM1}), $\psi $ is the (spinor) wave function of the quarks, $\gamma ^{\mu}$ are the Dirac matrices, $f_{abc}$ are the structure constants of the group SU(3), and $\alpha _s$ is the strong interaction coupling constant. In the first order perturbation theory, one can neglect the quark masses, so that the equation of
state for zero temperature quark matter can be obtained as \cite{Wein, Witt}
\begin{equation}\label{eq187}
p_q=\frac{1}{3}\left( \rho _qc^2-4B\right) ,
\end{equation}
where $B$ is interpreted physically as the difference between the energy density of the perturbative
and non-perturbative QCD vacua (the bag constant), while $\rho _q $ and $p_q$ denote the
energy density and thermodynamic pressure of the quark matter, respectively. Equation~(\ref{eq187}) is called the MIT bag model equation of state.
From a physical point of view, Eq.~(\ref{eq187}) represents the equation of state of a gas of massless particles (the quarks),
with corrections due to the QCD trace anomaly and perturbative interactions included.
The quarks are bound together by the vacuum pressure, $B$. Hence Eq.~(\ref{eq187}) provides a simplified
theoretical model for the long-range, quark confining interactions in QCD. The typical
value of the bag constant $B$, as obtained from particle physics experiments, and theoretical considerations, is of the order of $B=57\;{\rm MeV}/{\rm fm}^{3}\approx
10^{14}\;{\rm g}/{\rm cm}^{3}$ \cite{Wein,Witt}. On the other hand, after a neutron matter-quark matter phase transition,
which can take place, for example, in the dense core of neutron stars, the
energy density of strange quark matter is of the order of $\rho _q\approx 5\times 10^{14}\;{\rm g}/{\rm cm}^{3}$.

However, it is important to note that, in the case of the QCD description of strong interactions, the strong coupling constant $\alpha _s$ is a function of the particle (i.e. quark) momenta, and of their energy density. For the simplest hadronic models, the quark-gluon coupling constant is of the order of $\alpha _s \approx 0.12$ \cite{Wein}. If we define the generalised QCD charge as $Q_{QCD} \approx  \alpha _s^{1/2}$, we may obtain an estimate for the mass of a quark, interpreted as an electric and color-charged particle having a minimum mass/ratio, by applying the formalism developed in this paper and identifying the constant $\Lambda_*$ with $B$, where $B$ is the bag constant introduced in the simple MIT bag model, Eq.~(\ref{eq187}). This yields a value of order $m_q \approx 67.75$ MeV \cite{Boehmer:2007gq}, which represents a reasonable approximation to the predicted mass of the s quark \cite{Witt}.

\section*{Acknowledgments}

We would like to thank the anonymous referee for comments that helped us to improve our manuscript. P.B. and K.C. are supported, in part, by the Thailand Research Fund~(TRF), Commission on Higher Education~(CHE) and Chulalongkorn University under grant RSA5780002. M.L. is supported by a Naresuan University Research Fund individual research grant. T.H. and M.L. thank the Yat Sen School, the Department of Physics, and the Centre Sino-Fran{\c c}ais at Sun Yat Sen University, Guangzhou, P. R. China, for gracious hospitality during the final preparation of the manuscript.

\appendix

\label{appa}

\end{document}